\documentclass[11pt,draftclsnofoot,onecolumn,a4]{IEEEtran}

\usepackage{amsmath, mathrsfs}
\usepackage{amsfonts}
\usepackage{amssymb}
\usepackage{multirow}
\def\cop{{CPU}}

\usepackage[normalem]{ulem}

\usepackage{algorithm}
\usepackage{algpseudocode}
\usepackage{pifont}
\usepackage{varwidth}

\newcommand{\subparagraph}{}
\usepackage[compact]{titlesec}
\titlespacing*{\section}{0pt}{0.6\baselineskip}{0.5\baselineskip}
\titlespacing*{\subsection}{0pt}{0.4\baselineskip}{0.3\baselineskip}
\allowdisplaybreaks

\usepackage{caption}

\usepackage[numbers,sort&compress]{natbib}
\usepackage{balance}
\usepackage{multicol}


\usepackage{mathbbol}

\def\mindex#1{\index{#1}}

\newcommand{\blue}[1]{{\color{blue} #1}}

\def\sq{\hbox{\rlap{$\sqcap$}$\sqcup$}}
\def\qed{\ifmmode\sq\else{\unskip\nobreak\hfil
\penalty50\hskip1em\null\nobreak\hfil\sq
\parfillskip=0pt\finalhyphendemerits=0\endgraf}\fi\medskip}

\long\def\defbox#1{\framebox[.9\hsize][c]{\parbox{.85\hsize}{%
\parindent=0pt
\baselineskip=12pt plus .1pt      
\parskip=6pt plus 1.5pt minus 1pt 
 #1}}}

\long\def\beginbox#1\endbox{\subsection*{}%
\hbox{\hspace{.05\hsize}\defbox{\medskip#1\bigskip}}%
\subsection*{}}

\def\endbox{}

\newsavebox{\junk}
\savebox{\junk}[1.6mm]{\hbox{$|\!|\!|$}}

\def\argmin{\mathop{\rm arg\, min}}

\newcommand{\field}[1]{\mathbb{#1}}

\def\Re{\field{R}}

\def\ind{\field{I}}

\def\bA{{\mathbb A}}

\def\bC{{\mathbb C}}

\def\bE{{\mathbb E}}

\def\bI{{\mathbb I}}

\def\bR{{\mathbb R}}
\def\bS{{\mathbb S}}

\def\bX{{\mathbb X}}

\def\bZ{{\mathbb Z}}

\def\bba{{\mathbb a}}

\def\bbd{{\mathbb d}}

\def\bbh{{\mathbb h}}
\def\bbi{{\mathbb i}}

\def\bbp{{\mathbb p}}
\def\bbq{{\mathbb q}}

\def\bbx{{\mathbb x}}
\def\bby{{\mathbb y}}
\def\bbz{{\mathbb z}}

\def\bfA{{\bf A}}
\def\bfB{{\bf B}}
\def\bfC{{\bf C}}

\def\bfE{{\bf E}}

\def\bfH{{\bf H}}
\def\bfI{{\bf I}}

\def\bfK{{\bf K}}
\def\bfL{{\bf L}}
\def\bfM{{\bf M}}

\def\bfP{{\bf P}}
\def\bfQ{{\bf Q}}
\def\bfR{{\bf R}}
\def\bfS{{\bf S}}

\def\bfW{{\bf W}}
\def\bfX{{\bf X}}
\def\bfY{{\bf Y}}
\def\bfZ{{\bf Z}}

\def\bfa{{\bf a}}
\def\bfb{{\bf b}}

\def\bfe{{\bf e}}

\def\bfg{{\bf g}}
\def\bfh{{\bf h}}

\def\bfv{{\bf v}}

\def\bfx{{\bf x}}
\def\bfy{{\bf y}}
\def\bfz{{\bf z}}

\def\scrI{{\mathscr{I}}}

\def\scrL{{\mathscr{L}}}

\def\scrS{{\mathscr{S}}}

\def\ttF{{\mathtt{F}}}

\def\tta{{\mathtt a}}

\def\ttc{{\mathtt c}}
\def\ttd{{\mathtt d}}

\def\ttf{{\mathtt f}}

\def\tto{{\mathtt o}}
\def\ttp{{\mathtt p}}
\def\ttq{{\mathtt q}}

\def\ttt{{\mathtt t}}

\def\sfH{{\sf H}}

\def\sfw{{\sf w}}

\def\bfmath#1{{\mathchoice{\mbox{\boldmath$#1$}}%
{\mbox{\boldmath$#1$}}%
{\mbox{\boldmath$\scriptstyle#1$}}%
{\mbox{\boldmath$\scriptscriptstyle#1$}}}}

\def\bfmY{\bfmath{Y}}

\def\bfmhhaY{\bfmath{\hhaY}} 
\def\bfmhhaY{\hbox to 0pt{$\widehat{\bfmY}$\hss}\widehat{\phantom{\raise 1.25pt\hbox{$\bfmY$}}}}

\def\til={{\widetilde =}}

\def\clB{{\cal B}}
\def\clC{{\cal C}}
\def\clD{{\cal D}}

\def\clG{{\cal G}}

\def\clI{{\cal I}}
\def\clK{{\cal K}}

\def\clM{{\cal M}}
\def\clN{{\cal N}}

 \def\FRAC#1#2#3{\genfrac{}{}{}{#1}{#2}{#3}}

\def\ddtp{{\mathchoice{\FRAC{1}{d^{\hbox to 2pt{\rm\tiny +\hss}}}{dt}}%
{\FRAC{1}{d^{\hbox to 2pt{\rm\tiny +\hss}}}{dt}}%
{\FRAC{3}{d^{\hbox to 2pt{\rm\tiny +\hss}}}{dt}}%
{\FRAC{3}{d^{\hbox to 2pt{\rm\tiny +\hss}}}{dt}}}}

\def\eqdef{\mathbin{:=}}

\def\average#1,#2,{{1\over #2} \sum_{#1}^{#2}}

\def\eye(#1){{\bf(#1)}\quad}

\newtheorem{theorem}{{\bf Theorem}}

\newtheorem{remark}{{\bf Remark}}

\newtheorem{proposition}[theorem]{{\bf Proposition}}

\def\eq#1/{(\ref{e:#1})}

\newcommand{\inp}[2]{{\langle #1, #2 \rangle}}
\newcommand{\inpr}[2]{{\langle #1, #2 \rangle}_\bR}

\newcommand{\beqn}[1]{\notes{#1}%
\begin{eqnarray} \elabel{#1}}

\newcommand{\eeqn}{\end{eqnarray} }

\newcommand{\beq}[1]{\notes{#1}%
\begin{equation}\elabel{#1}}

\newcommand{\eeq}{\end{equation}}

\def\bdes{\begin{description}}
\def\edes{\end{description}}

\newcounter{rmnum}

\newcounter{anum}

{\end{list}}

\def\ass(#1:#2){(#1\ref{#1:#2})}

\def\ritem#1{
\item[{\sf \ass(\current_model:#1)}]
}

\newenvironment{recall-ass}[1]{%
\begin{description}
\def\current_model{#1}}{
\end{description}
}

\usepackage[pdftex]{graphicx}
\usepackage{subfig,float}
\usepackage{color}
\usepackage{tikz}
\usepackage{pgfplots}
\pgfplotsset{compat=newest}
\usetikzlibrary{patterns}
\usetikzlibrary{positioning}
\usetikzlibrary{datavisualization}
\usetikzlibrary{datavisualization.formats.functions}
\usetikzlibrary{backgrounds}
\usetikzlibrary{shapes,snakes}

\DeclareGraphicsExtensions{.eps,.pdf,.png,.jpg,.gif,.jpeg,.pstex}
\def\gammas{{\gamma_{\ttd}}}
\def\gammai{{\gamma_{\ttc\tto}}}

\usepackage[none]{hyphenat}

\def\herm{{\sfH}}

\def\sinr{{\mathsf{sinr}}}

\newcommand{\range}[2]{{\text{$#1$\,:\,$#2$}}}

\def\cg{{\clC\clN}}

\def\prox{{\mathsf{prox}}}

\def\interF{{\bfE}}
\def\interf{{\bfe}}

\def\matlab{{MATLAB\textcopyright\,}}
\def\clst{{\mathsf{Clst}}}
\def\fovs{{\ttF^{\text{ovs}}}}
\def\ford{{\ttF}}

\usepackage{xcolor}

\long\def\comment#1{}

\newcommand{\av}{{\bf a}}

\newcommand{\hv}{{\bf h}}

\newcommand{\xv}{{\bf x}}

\newcommand{\Hm}{{\bf H}}

\newcommand{\Xm}{{\bf X}}

\newcommand{\Zm}{{\bf Z}}

\newcommand{\Kc}{{\cal K}}

\newcommand{\Xc}{{\cal X}}

\newcommand{\Lambdam}{\hbox{\boldmath$\Lambda$}}

\newcommand{\Sigmam}{\hbox{\boldmath$\Sigma$}}

\newcommand{\trace}{{\hbox{tr}}}

\renewcommand{\Re}{{\rm Re}}

\newcommand{\transp}{{\sf T}}
\renewcommand{\vec}{{\rm vec}}

\begin{document}

\title{Massive MIMO Pilot Decontamination and Channel Interpolation via Wideband Sparse Channel Estimation}
\author{Saeid Haghighatshoar,  \IEEEmembership{Member, IEEE,} Giuseppe Caire,
\IEEEmembership{Fellow, IEEE}%
\thanks{The authors are with the Communications and Information Theory Group, Technische Universit\"{a}t Berlin (\{saeid.haghighatshoar, caire\}@tu-berlin.de).

A short version of this paper was presented in 50\,th Annual Asilomar Conference on Signals, Systems, and Computers (Asilomar 2016).
}
\vspace{-0.5cm}
}

\maketitle

\begin{abstract}
We consider a massive MIMO system based on \textit{Time Division Duplexing} (TDD) and channel reciprocity, 
where the base stations (BSs) learn the channel vectors of their users via the pilots transmitted by the users in the \textit{uplink} (UL). 
It is well-known that, in the limit of very large number of BS antennas, the system performance is limited by pilot contamination, due to the fact that 
the same set of orthogonal pilots is reused in multiple cells.  In the regime of moderately large number of antennas, 
another source of degradation is channel interpolation because the pilot signal of each user
probes only a limited number of OFDM subcarriers and the channel must be interpolated over the other subcarriers where no pilot symbol is transmitted. 
In this paper, we propose a low-complexity algorithm that uses the received UL wideband pilot snapshots 
in an observation window comprising several coherence blocks (CBs) to obtain an estimate of the \textit{angle-delay Power Spread Function} (PSF) 
of the received signal. This is generally given by the sum of the angle-delay PSF of the desired user and the angle-delay PSFs of the copilot users (\cop s), i.e., the users re-using the same pilot dimensions in other cells/sectors. 
We propose supervised and unsupervised clustering algorithms to decompose the estimated PSF and isolate the part corresponding to  
the desired user only.  We use this decomposition to obtain an estimate of the  covariance matrix of the user wideband channel vector, 
which we exploit to decontaminate the desired user channel estimate by applying \textit{Minimum Mean Squared Error} (MMSE) smoothing filter, i.e., the optimal
channel interpolator in the MMSE sense. 
We also propose an effective low-complexity approximation/implementation of this smoothing filter. 
We use numerical simulations to assess the performance of our proposed method, and compare it with other recently proposed schemes that use the same idea of separability of users in the angle-delay domain.
\end{abstract}


\section{Introduction}  \label{intro}
{
Consider a massive MIMO multi-cell system with $M$ antenna per each base station (BS), per-cell processing,  
\textit{Orthogonal Frequency Division Multiplexing} (OFDM), 
\textit{Time Division Duplexing} (TDD), and reciprocity-based channel estimation
as in \cite{Marzetta-TWC10,shepard2012argos,larsson2014massive}. 
In such systems, time is divided into several slots, where in each slot users are scheduled to send {\em uplink} (UL) pilot signals in order to allow the BS to estimate their channel vectors. The BS exploits the UL-DL reciprocity and uses the resulting  channel estimates  to coherently detect data from the users in the UL and precode data to the users in the DL. 
A family of mutually orthogonal pilot sequences are obtained in the time-frequency domain by assigning to each pilot a different set of 
signal dimensions in the tessellation of the time-frequency plane under the OFDM \cite{Marzetta-TWC10} or, more in general, 
by sharing all the signal dimensions but assigning  to the pilots mutually orthogonal symbol sequences across all signal dimensions (e.g., see \cite{you2016channel}). 
Due to limited channel coherence time, the signal dimensions in each UL-DL scheduling slot are limited. Consequently, 
also the number of UL pilot signal dimensions is limited, resulting in a limited number of orthogonal pilots.  Therefore, to simultaneously serve several users across the whole system, pilots must be reused in multiple cells according 
to a specific reuse pattern \cite{Marzetta-TWC10}. As a result, the channel estimation during the UL pilot transmission is severely degraded by the interference received from the users in 
neighboring cells (or sectors) re-using the same pilot sequences as the users inside the cell; these users are referred to as copilot users (\cop s).
Such a phenomenon is called {\em pilot contamination}. It is well-known that  
pilot contamination becomes the only limiting factor on the spectral efficiency of the system in the asymptotic limit where the number of BS antennas $M \rightarrow \infty$ 
but the number of users per cell $K$ is kept finite \cite{Marzetta-TWC10,jose2011pilot}. In the more realistic case of large but finite $M$ and $K$ with $M \gg K$,  the pilot contamination still represents an important source of degradation especially for the edge users lying on the cell boundary \cite{Huh11,hoydis2013massive,bjornson2016massive}.
}

\subsection{Approaches to pilot decontamination} 
{Several approaches have been proposed to cope with pilot contamination. 
In \cite{yin2013coordinated}, it is observed that if multipath components (MPCs) of the channel vectors of the users have a limited 
angular spread (spatial correlation), it is possible to coordinate the pilot transmission in adjacent cells such that the channels of \cop s are confined in
nearly orthogonal subspaces due to their angular diversity. However, in order to effectively separate \cop s, the covariance information (or subspace information), i.e., the second-order statistics of users' channel vectors, must be known at the BS. A similar a priori statistical knowledge is used in 
\cite{adhikary2013joint,nam2014joint} in the so-called JSDM scheme to reuse pilots in the same cell in order to decrease the pilot dimension 
overhead. More generally,  it has been shown  in \cite{bjornson2016pilot} that if  the covariance matrices of the users and their \cop s are available 
at the BS and satisfy certain mild conditions of linear independence, pilot contamination in the limit of $M \rightarrow \infty$ can be completely 
eliminated. However, this requires the knowledge of the user channel covariance matrices,  which is itself difficult to obtain precisely due to 
 pilot contamination.

A quite different approach is proposed in \cite{muller2014blind}, in which no a priori knowledge of subspace is needed. 
Instead, it is noticed that when the number of BS antennas $M$ is much larger than the number of per-cell served users $K$, 
and the power imbalance between the desired and the interfering users is above a certain threshold, the eigenvalues of the sample covariance matrix of the received signal corresponding to the desired users and that corresponding to \cop s in adjacent cells concentrate on ``clusters'' with disjoint supports. Thus, by distinguishing those clusters, it is possible to identify blindly the desired and the interfering signal subspaces.
In contrast to \cite{yin2013coordinated,adhikary2013joint,nam2014joint}, which work only in the presence of spatially correlated channels, 
the method of \cite{muller2014blind} would work also with i.i.d. (isotropically distributed) channel vectors, provided that the power imbalance between 
the desired and the interfering users is sufficiently large and the matrix dimension is large enough such that the eigenvalue clustering is sufficiently sharp. 
A combination of techniques in \cite{yin2013coordinated} and \cite{muller2014blind}, via exploiting both the spatial correlation and power discrimination, has been used in \cite{yin2016robust}.

Another method to cope with pilot contamination consists in ``pilot contamination precoding'' as proposed in \cite{li2013pilot}. 
The main idea is that, due to very large number of BS antennas, the only residual interference that matters after beamforming is the coherent interference due to pilot contamination, which can be eliminated by jointly precoding across neighboring cells (e.g., in the DL using linear precoding or non-linear dirty-paper coding and in the UL using linear interference mitigation or non-linear successive interference cancellation). Such a scheme, however, requires centralized processing of multiple cell sites in order to jointly decode/precode the UL/DL signals; this goes against the beauty and simplicity of massive MIMO, for which single-cell processing is one of the main motivations  \cite{Marzetta-TWC10}. 

In the recent work \cite{chen2016pilot}, developed independently and in parallel with our present work, 
a method for pilot decontamination is proposed by exploiting the fact that the channel vectors of \cop s at a given BS have typically different 
MPCs in the angle-delay domain.  Therefore, if it is possible to identify the MPCs pertaining only to the desired user, the interference due to \cop s can be mitigated by linear  space-frequency filtering, thus, mitigating the effect of 
pilot contamination. 
Our work is also based on the same idea but differs from \cite{chen2016pilot} in many aspects and generally can achieve much better performance 
without incurring any additional pilot overhead with respect to the standard pilot schemes used in current systems (e.g., in LTE-TDD \cite{holma2011lte}). 
We defer a through comparison of \cite{chen2016pilot} with our work to Section \ref{sec:performance}. 
}

\subsection{Contribution}

In this paper, we pursue a new method for pilot decontamination that has the following advantages:
\begin{itemize}
\item Unlike \cite{yin2013coordinated, yin2016robust, bjornson2016pilot}, we do not assume a priori knowledge of the channel covariance matrices or centralized coordination of pilot allocation. 
\item Unlike \cite{muller2014blind, yin2016robust}, we do not rely on asymptotic results in random matrix theory, which requires 
a)\,i.i.d. isotropic channel vectors (spatially correlated channel vectors along with covariance information in \cite{yin2016robust}), and b)\,sufficiently large power imbalance between the users inside and outside the cell.
\item Unlike \cite{li2013pilot}, we do not rely on joint precoding and centralized processing. Instead, we apply a strictly uncoordinated per-BS processing. 
\end{itemize}
\begin{figure}[t]
\centering
\includegraphics[width=7.5cm]{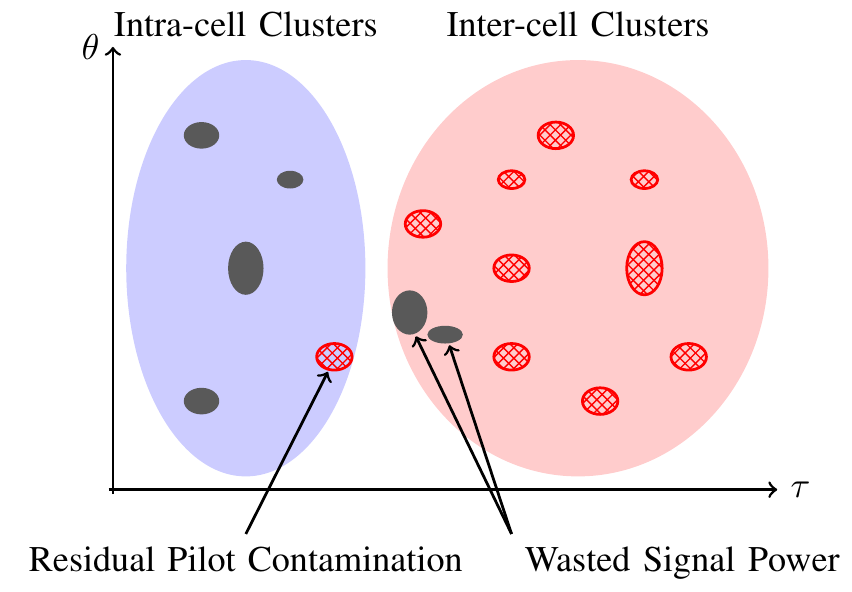}
\caption{\small Illustration of the proposed pilot decontamination scheme via exploiting the sparsity of  the angle-delay scattering map of the received signal at the BS. The multipath components (MPCs) of the intended user and those of its \cop s are illustrated with grey solid and red dashed bubbles respectively. In this example, most of the MPCs of the user have  shorter delays, thus, they can be identified and fairly separated from those of the \cop s via clustering in the delay domain as illustrated in the figure. The red dashed bubbles inside the signal cluster illustrate the residual copilot interference whereas the grey solid bubbles inside the interference cluster correspond to the useful signal wasted because of the delay-based clustering.}
\label{multicell}
\end{figure}
Here, we only provide an intuitive explanation of our proposed scheme  and postpone the thorough description to  Section \ref{sec:PDC}. 
The idea is qualitatively illustrated in Fig.\,\ref{multicell}. {In a massive MIMO macrocell system, the propagation between users and BS antennas 
occurs through relatively \textit{sparse} MPCs in the angle-delay domain. We exploit this underlying sparsity to estimate the \textit{angle-delay Power Spread Function} (PSF) of each user by sampling only a small number of antennas and sending UL pilots only over a small subset of subcarriers. 
Then, we apply suitable algorithms to cluster the estimated PSF in the angle-delay plane to approximately 
separate the MPCs belonging to the desired user from those of its \cop s. This is illustrated qualitatively in 
Fig.\,\ref{multicell} for a configuration where clustering is based on the difference of propagation delays, and where 
the interference due to \cop s can be fairly eliminated by filtering in the delay domain, at the cost of possibly filtering out also some components 
of the useful signal.} 
Furthermore, once we identified the angle-delay domain clusters pertaining to the desired user's PSF, we exploit them to obtain a very compact representation of the user wideband covariance matrix over the whole set of OFDM subcarriers. 
In turns, we use this information for MMSE channel estimation, obtaining at once both decontamination (i.e., the contribution of the \cop s
is filtered out by the channel estimator) and channel interpolation over the whole signal bandwidth.  
We develop a novel computationally efficient channel interpolation method that approximates the \textit{Minimum Mean Squared Error} (MMSE) 
smoothing filter. This provides a close-to-optimal MSE channel estimator under the Gaussian statistics and avoids performance degradation incurred due to imperfect instantaneous channel estimation, especially for a moderate number of antennas $M$ \cite{shirani2009channel}. 
%
%
%
%

\subsection{Notation}

We represent scalar constants by non-boldface letters (e.g., $x$ or $X$), sets by calligraphic letters (e.g., $\Xc$),  vectors by boldface small letters (e.g., $\xv$), and matrices by boldface capital letters (e.g., $\Xm$). 
We denote the $i$-th row and the $j$-th column of a matrix $\bfX$ with the row-vector $\Xm_{i,.}$ and the column-vector $\Xm_{.,j}$ respectively.
For a $p\times q$ matrix $\bfX$, we represent by $\vec(\bfX)$ the $pq\times 1$ column-vector obtained by stacking the column of $\bfX$ on top of each other, where we denote  the resulting vector with a blackboard letter $\bbx$ and a matrix consisting of $r$ such vectors by $\bX=[\bbx_1, \dots, \bbx_r]$.
We indicate the Hermitian conjugate and the transpose of a matrix  $\bfX$ by $\bfX^\herm$ and $\Xm^\transp$ with the 
same notation being used for vectors and scalars. $\bfX \otimes \bfY$ indicates the Kronecker product of the matrices $\bfX$ and $\bfY$. 
We denote the complex and the real inner product between two matrices (and similarly two vectors) $\bfX$ and $\bfY$ 
by $\inp{\bfX}{\bfY}=\trace(\bfX^\herm \bfY)$ and $\inpr{\bfX}{\bfY}=\Re[\inp{\bfX}{\bfY}]$ respectively. We use $\|\bfX\|=\inp{\bfX}{\bfX}$ for the Frobenius norm of a matrix $\bfX$ and $\|\bfx\|$ for the $l_2$-norm of a vector $\bfx$.
An  identity matrix of order $p$ is represented by $\bfI_p$. For an integer $k>0$, we use the shorthand notation $[k]$ for $\{1,2,\dots, k\}$. 
%

{
\section{Problem Statement}\label{sec:pilot-contamination}
\subsection{Basic Setup} 
Our model and system assumptions are standard  in most classical works on massive MIMO
(e.g.,  \cite{Marzetta-TWC10,shepard2012argos,larsson2014massive,you2016channel,Huh11,hoydis2013massive,bjornson2016massive,
adhikary2013joint,nam2014joint,bjornson2016pilot,yin2013coordinated,chen2016pilot}) and recalled here for the sake of completeness and for 
establishing the notation to be used later. 
We consider a system with a signal bandwidth of $W$\,Hz and a scheduling slot of duration $T_s$\,sec (including UL pilots, UL payload, and DL payload \cite{Marzetta-TWC10}). 
The underlying channel fading process has a coherence bandwidth $\Delta f_c < W$ 
and a coherence time $\Delta t_c \geq T_s$ \cite{tse2005fundamentals}, such that in each scheduling slot 
we have $\lceil \frac{W}{\Delta f_c} \rceil$  frequency sub-bands over which the channel can be considered (approximately) frequency-flat
and constant in time over the whole duration of a slot. We call a frequency-time rectangle of bandwidth $\Delta f_c$ and duration $T_s$ a \textit{coherence block} (CB). This is illustrated in Fig.\,\ref{pilot_structure}, where it is seen that the channel is approximately constant over a CB but changes smoothly across different CBs. We denote by $\Delta \tau_{\max}$ the maximum channel delay spread that the system can handle without suffering from inter-block interference between the OFDM symbols \cite{molisch2012wireless}. We assume that a set of $B$ OFDM symbols are transmitted inside a time slot, each having a total duration of $T_\text{OFDM}=\frac{T_s}{B}$ and an effective duration of $T_u=T_\text{OFDM}-\Delta \tau_{\max}$ after removing the \textit{cyclic prefix} (CP) of duration $\Delta \tau_{\max}$. The frequency spacing between the subcarriers is given by $\Delta f=\frac{1}{T_u}$, thus, each OFDM symbol has $N=\frac{W}{\Delta f}=W T_u$ subcarriers. Over each slot, we have a set of $Q=N B=W T_u B = W T_s (1- \frac{\Delta \tau_{\max}}{T_\text{OFDM}})$
signal dimensions. 
\begin{figure}[t]
\centering
\includegraphics[width=10cm]{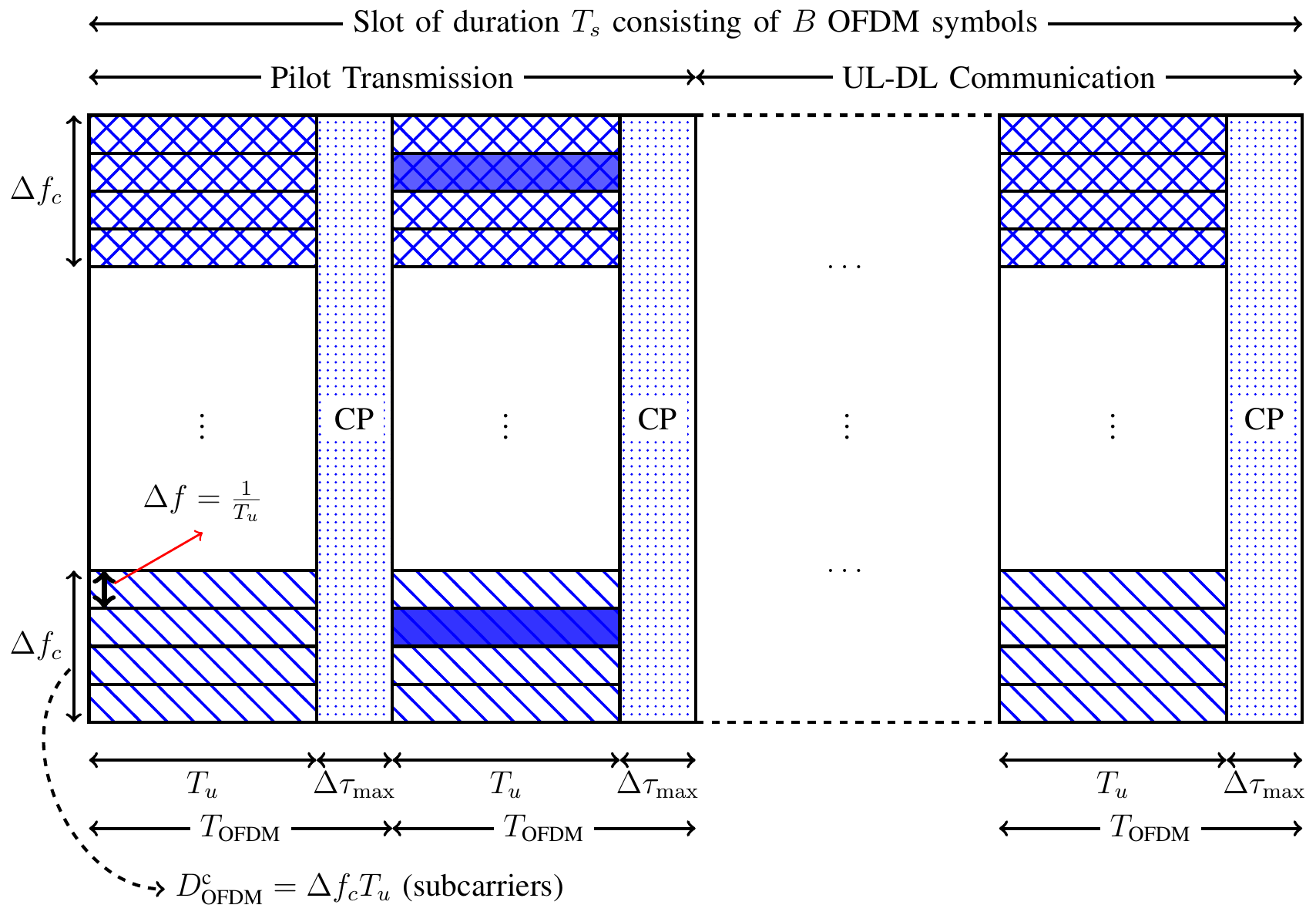}
\caption{A slot of duration $T_s$ consisting of several OFDM symbols. In this example, a coherence sub-block inside an OFDM symbol contains $D^\text{c}_\text{OFDM}=4$ signal dimensions (subcarriers), and two OFDM symbols are devoted to pilot transmission, thus, $D^\text{c}_\text{p}=2D^\text{c}_\text{OFDM}=8$. We also illustrate a 0-1 pilot sequence that lies on the second training OFDM symbol  and probes subcarrier \#3 inside each coherence sub-block.}
\label{pilot_structure}
\end{figure}
Also, each CB is decomposed into $B$ disjoint sub-blocks lying inside separate OFDM symbols, where each sub-block consists of $D^\text{c}_\text{OFDM}=\Delta f_c T_u$  subcarriers and, in total, there are $D^\text{c}= B D^\text{c}_\text{OFDM}$ signal dimensions in each CB. During each slot,  some $D^\text{c}_p$ out of $D^\text{c}$ signal dimensions inside each CP are devoted to pilot transmission, while the remaining $D^\text{c}_d=D^\text{c}-D^\text{c}_p$ signal dimensions are used for UL-DL data transmission. A set of $D^\text{c}_p$ orthogonal pilot sequences are assigned to $D^\text{c}_p$ pilot signal dimensions in each CB. The resulting orthogonal pilots are allocated to the users in each cell/sector according to 
a given reuse pattern (e.g., see \cite{Marzetta-TWC10,Huh11}), where in reuse patterns with a reuse factor $r\geq 1$, at most $K = \frac{D^\text{c}_p}{r}$ users can be simultaneously served per cell/sector with mutually orthogonal pilot sequences.
In this paper, without any loss of generality, we consider 0-1 pilot sequences (see Fig.\,\ref{pilot_structure}), where the pilot sequence of each user is transmitted over a single OFDM symbol and places a single ``1'' in each CB. In this way, each pilot sequence probes one subcarrier per CB (see, e.g., \cite{you2016channel}), for a total of 
$n=\frac{N}{D^\text{c}_\text{OFDM}}$ subcarriers. 
}

\subsection{Pilot Contamination}\label{sec:pilot_cont_new}
{We consider a reference BS called BS$_0$ and denote by UE$_{0,k}$ a generic user $k\in [K]$ served by BS$_0$. As before, we assume that the pilot signal of UE$_{0,k}$ is transmitted over an individual OFDM symbol and probes a subset of subcarriers $\Omega_k \subset [N]$ of size $|\Omega_k|=n$. We denote by $\clK_k$ the set of all \cop s of UE$_{0,k}$, i.e.,  the users across the whole system that transmit their pilot signal over the same pilot OFDM symbol and over the same set of subcarriers $\Omega_k$ as   UE$_{0,k}$. 
The received signal of UE$_{0,k}$ at BS$_0$ during the pilot transmission is given by
\begin{align}\label{pilot_cont_new}
\bfy_{k,s}[\omega]= \bfh_{k,s}[\omega] + \sum_{k' \in \clK_k} \bfh_{k',s}[\omega] + \bfz_s[\omega], \omega \in \Omega_k,
\end{align}
where $\bfh_{k,s}[\omega]$ and $\bfh_{k',s}[\omega]$ denote the $M$-dim channel vectors of UE$_{0,k}$ and its \cop s  $\clK_k$ to the $M$ antennas at BS$_0$ at time slot $s$ and subcarrier $\omega$, where $\bfz_s[\omega]\sim \cg(0, \sigma^2 \bfI_M)$ is the \textit{additive white Gaussian noise} (AWGN) at subcarrier $\omega$, and where we assumed, without loss of generality, that the transmitted pilot symbols at all subcarriers $\omega\in \Omega_k$ are normalized to $1$. From \eqref{pilot_cont_new}, it is seen that during the UL pilot transmission phase, the BS receives the superposition of the channel vector of UE$_{0,k}$ and that of its \cop s, thus, pilot contamination.
}
\subsection{Wideband Pilot Decontamination}
{
We denote by $\Hm_{k,s} = \big[\hv_{k,s}[1], \dots , \hv_{k,s}[N]\big]$ and $\Hm_{k',s} = \big[\hv_{k',s}[1], \dots , \hv_{k',s}[N]\big]$, $k'\in \clK_k$, the $M\times N$ wideband channel matrices of UE$_{0,k}$ and its \cop s across $N$ OFDM subcarriers at time slot $s$.  We denote by $\bfS^\ttf _{k,s}$ an $n\times N$ matrix  that has a single $1$ in each row at columns corresponding to the probed subcarriers $\Omega_k$ and is $0$ elsewhere. We also assume, for the sake of generality,  
that during the UL training phase,  a subset  of size $m$ of the $M$ BS antennas is sampled via an $m\times M$ matrix $\bfS^\tta_s$. 
Thus, from \eqref{pilot_cont_new}, the UL pilot observation for UE$_{0,k}$ at BS$_0$ at time slot $s$ can be arranged as a $m \times n$ matrix 
\begin{equation} \label{UL-pilot-measurement}
\Xm_{k,s}:= \bfS^{\tta}_{s} \bfY_{s} {\bfS^{\ttf}_{k,s}}^{\herm}= \widetilde{\Hm}_{k,s} +  \sum_{k' \in \Kc_k} \widetilde{\Hm}_{k',s} + \widetilde{\Zm}_{k,s}, 
\end{equation} 
where $\bfY_s=\big[\bfy_s[1], \dots, \bfy_s[N]\big ]$ denotes the $M\times N$ wideband signal received across all the subcarriers, and where $\widetilde{\Hm}_{k,s}=\bfS^{\tta}_{s} \bfH_{k,s} {\bfS^{\ttf}_{k,s}}^{\herm}$ contains the channel coefficients of UE$_{0,k}$ corresponding to the 
$m$ sampled antennas and the $n$ probed subcarriers, with the same interpretation holding for $\widetilde{\Hm}_{k',s}$, $k' \in \clK_k$, and $\widetilde{\Zm}_{k,s}$. 
We denote by $\bbh_{k,s}=\vec(\bfH_{k,s})$ and $\bbh_{k',s}=\vec(\bfH_{k',s})$, $k'\in \clK_k$, the \textit{wideband channel vector}s obtained after vectorization. Applying the $\vec$ operator and using the identity $\vec(\bfA \bfB \bfC)=(\bfC^\transp \otimes \bfA) \vec (\bfB)$, we can write \eqref{UL-pilot-measurement} as 
\begin{align}\label{intro_wideband_sketch}
\bbx_{k,s}=\bS_{k,s} \bby_{k,s}= \widetilde{\bbh}_{k,s} + \sum_{k'\in \clK_k} \widetilde{\bbh}_{k',s} + \widetilde{\bbz}_{k,s},
\end{align}
where $\bS_{k,s}=\bfS^\ttf _{k,s} \otimes \bfS^\tta_s$ and where $\widetilde{\bbh}_{k,s}=\bS_{k,s} \bbh_{k,s}$, $\widetilde{\bbh}_{k',s}=\bS_{k',s} \bbh_{k',s}$, $k'\in \clK_k$. With this notation, the objective of pilot decontamination can be stated as follows.
\vspace{3mm}

\noindent
\colorbox{gray!40}{ 
\begin{minipage}{0.97\textwidth}
\noindent{\bf Pilot Decontamination:} 
Given the noisy and contaminated UL \textit{wideband} pilot sketches $\{\bbx_{k,s} : s \in [\sfw]\}$ of the desired user   UE$_{0,k}$ across $\sfw$ time slots, construct an estimator for its wideband channel vector $\bbh_{k,s}$ (equivalently, its wideband channel matrix $\bfH_{k,s}$) at the next time slots $s \geq \sfw + 1$.  
$\phantom{\sum}$  \hfill $\lozenge$
\end{minipage}
}

\vspace{3mm}

\noindent To explain this better, let us define the \textit{wideband} (\textit{space-frequency}) covariance matrices of UE$_{0,k}$ and of its \cop s by $\bfC_{k} = \bE[\bbh_{k,s}  \bbh_{k,s}^\herm]$ and $\bfC_{k'}=\bE[\bbh_{k',s}  \bbh_{k',s}^\herm]$, $k' \in \Kc_k$, independent of $s$ 
by the WSS assumption (see Section \ref{sec:wss}).
Note that if these covariance matrices are available at BS$_0$, using the fact that the channel vectors $\bbh_{k,s}$ and $\{\bbh_{k',s}: k' \in \clK_k\}$ are 
independent vector-valued stationary Gaussian  random processes (see Section \ref{sec:wss} for more details), the immediate answer to  our estimation problem for pilot decontamination would be the MMSE smoothing filter, given by\footnote{Notice that here, knowing the \textit{space-frequency} covariance matrices, we used
only the observation at slot $s$ to estimate the channel at slot $s$. In general, we can use $\bbx_{k,s}$ together with the all the past observations $\{\bbx_{k,s'}: s' <s\}$ to do pilot decontamination, but this will require estimating the \textit{space-frequency-doppler} covariance matrices of the current and past observations, which would result in even a more complex estimator. In practice, since the slot time $T_s$ is usually chosen to be 
of the same order of the channel coherence time $\Delta t_c$, the channel samples at different slots are nearly independent, and there is very little to gain from the temporal correlation of the fading. For this reason, we restrict to the common practice of estimating the channel 
based on the current slot UL pilot observation \cite{Marzetta-TWC10,shepard2012argos,larsson2014massive}.}
\begin{equation} \label{MMSE-smoothing}
\widehat{\bbh}_{k,s} =\Sigmam_{\bbh_k, \bbx_k} \bfC_{\bbx_k}^{-1} \bbx_{k,s} = \bfC_{k} \bS^\herm_{k,s} \left (\sigma^2 \bfI_{mn} + \bS_{k,s} \left ( \bfC_{k} + \sum_{k' \in \Kc_k} \bfC_{k'} \right ) \bS^\herm_{k,s} \right )^{-1} \bbx_{k,s},
\end{equation}
where $\Sigmam_{\bbh_k,\bbx_k}:=\bE[\bbh_{k,s} \bbx_{k,s}^\herm]=\bfC_{k} \bS^\herm_{k,s}$, and where we used $\bS_{k,s}^\herm \bS_{k,s}=\bfI_{mn}$. 
In practice, however, $\bfC_k$ and $\bfC_{k'}$, $k'\in \clK_k$, are not available and should be estimated from the noisy and contaminated pilot sketches $\{\bbx_{k,s}: s\in [\sfw]\}$. With this brief explanation, the problems we are addressing in this paper are as follows:
\begin{enumerate}
\item \textit{How can we efficiently estimate the wideband covariance matrices
of the desired user and of the \cop s from the subsampled and contaminated observations $\{\bbx_{k,s} : s \in [\sfw]\}$?} We address this question by estimating the contaminated wideband channel covariance matrix $\bfC_k + \sum_{k'\in \clK_k} \bfC_{k'}$  via exploiting the sparsity of MPCs in the angle-delay domain  (Section \ref{sec:ch_estim}),  and applying suitable clustering techniques in the angle-delay domain to decompose approximately the resulting contaminated wideband covariance matrix into its signal and interference parts $\bfC_k$ and $\sum_{k'\in \clK_k} \bfC_{k'}$ (Section \ref{sec:cluster} and \ref{PDC_opt}). 

\item \textit{How can we approximate the MMSE smoothing filter (\ref{MMSE-smoothing}) in an efficient way, not requiring inversion of a $mn \times mn$ matrix  and complicated and time-consuming matrix-matrix multiplication in (\ref{MMSE-smoothing})?} We address this complexity issue by developing computationally-efficient pilot decontamination and channel interpolation  algorithms (Section \ref{PDC_subopt} and Appendix \ref{sec:admm}).
\end{enumerate}


}

\section{Wideband Channel Model}  \label{sec:directional}

\subsection{WSS-US Assumption}\label{sec:wss} 

\begin{figure}[ht]
\centerline{\includegraphics[width=8cm]{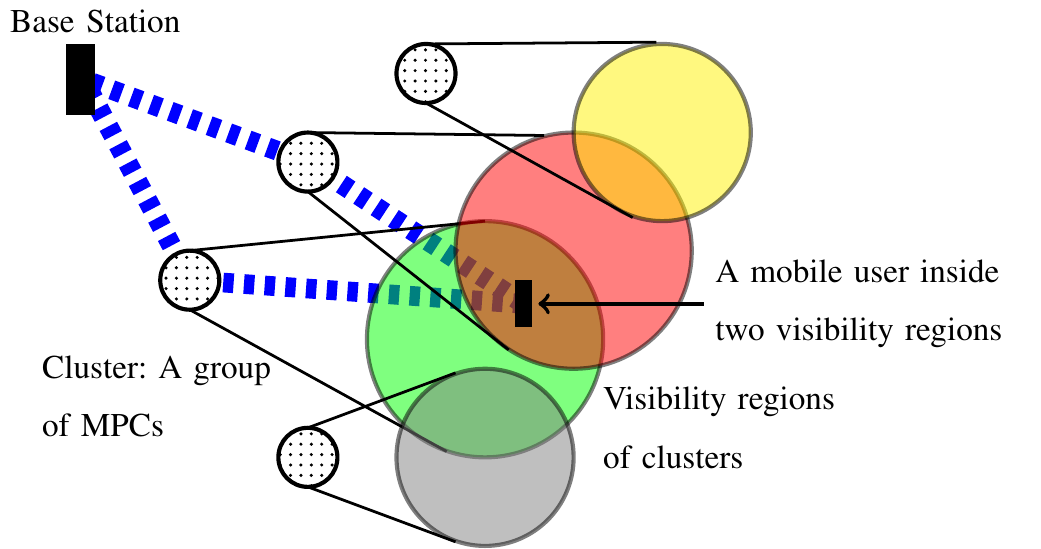}}
\caption{A sketch of the cluster and visibility regions of the COST2100 model.}
\label{COST2100}
\end{figure}

The COST 2100 channel model consists (up to some drastic simplifications) of clusters of MPCs and visibility regions \cite{liu2012cost}. 
The propagation between a BS and a user inside the intersection of multiple visibility regions occurs through all 
corresponding clusters (see Fig.\,\ref{COST2100}). 
This implies that the statistics of the channel between a BS and a user remains constant in time
and frequency as long as the user remains in the intersection of same visibility regions. As the user crosses the boundary of some region and enters a new region, the channel statistics typically undergoes a sharp transition. Since moving across the regions occurs at a time scale much larger than
moving across one wavelength, it is safe to assume that the channel statistics is piecewise time-invariant with 
relatively sharp transitions at very low rate compared with the signaling rate.  
In this paper, for simplicity, we neglect such transitions and suppose a time-invariant second-order statistics for the channel during the whole communication interval, i.e., the channel process is assumed to remain (locally) \textit{Wide Sense Stationary} (WSS) over time.  
Furthermore, the MPCs originated by different users and/or different scattering clusters are assumed to be mutually uncorrelated (US assumption). 
Finally, since each MPC is formed by a very large number of elementary multipath contributions, superimposing with different phases, 
invoking the Central Limit Theorem it is widely accepted to model the MPC coefficients as  complex circularly symmetric Gaussian \cite{tse2005fundamentals,molisch2012wireless}. 

\subsection{Sparsity in the Angle-Delay Domain}\label{sec:wide_channel}
Without loss of generality, we focus on a single BS-user pair and neglect the user and BS indices to simplify the notation. 
Also, for simplicity, we adopt a  discrete multipath model \blue{\cite{clerckx2013mimo,liu2003capacity,auer20123d,fleury2000first,you2016channel}} 
with $L$ MPCs, each of which is characterized by an \textit{Angle of Arrival} (AoA) $\theta_l$ and a delay $\tau_l$.   
All the results of this paper extend to the general case of mixed-type discrete-continuous 
scattering as long as the MPCs have a limited angle-delay support. 
In each time slot $s$, the UL channel is given by the vector impulse response 
\begin{equation} \label{vector-channel-impulse}
\check{\hv}_s(\tau) = \sum_{l=1}^L \rho_{s,l} \av(\theta_l) \delta(\tau - \tau_l),
\end{equation}
where $\bfa(\theta)$ denotes the array response at AoA $\theta \in [-\theta_{\max}, \theta_{\max}]$, 
whose $k$-th component is $[\bfa(\theta)]_k=e^{jk\frac{2\pi d}{\lambda} \sin(\theta)}$, 
where $\lambda=\frac{c_0}{f_0}$ denotes the wavelength ($c_0$ denoting  the speed of light) 
and where $f_0$ is the carrier frequency. We assume that the array elements have the uniform spacing $d=\frac{\lambda}{2\sin(\theta_{\max})}$, thus, 
$[\bfa(\theta)]_k= e^{jk\pi \frac{\sin(\theta)}{\sin(\theta_{\max})}}$. 
As said before, from the WSS-US and Gaussian assumption, we have that 
the discrete-time path gain processes $\{\rho_{s,l} : s \in \bZ\}$ are stationary with respect to the (slot) time index $s$ 
and independent across $l$. Furthermore, we assume no line of sight propagation, yielding 
$\rho_{s,l} \sim \cg(0, \sigma_{l}^2)$, where $\sigma_l^2$ denotes the  strength of the $l$-th MPC (independent of $s$ because of the WSS assumption). 

In the OFDM discrete frequency domain, channel (vector) frequency response corresponding to 
the impulse response \eqref{vector-channel-impulse} is given by 
\begin{eqnarray} \label{ziocanale2}
\hv_s[\omega] =  \sum_{l = 1}^L \rho_{s,l} \av(\theta_l) e^{-j \frac{2\pi}{N} W \tau_l (\omega - 1)}, \ \omega \in [N],  
\end{eqnarray} 
such that, as anticipated in Section \ref{sec:pilot-contamination}, the wideband channel matrix at slot $s$ is given by
\begin{align}\label{H_s}
\bfH_s=\Big [\bfh_s[1], \bfh_s[2], \dots, \bfh_s[N] \Big ].
\end{align}
We define the $N$-dim vector $\bfb(\tau)$, whose $\omega$-th component given by
$[\bfb(\tau)]_\omega = e^{j \frac{2\pi}{N} W \tau (\omega - 1)}$. Thus, we can write \eqref{H_s} more compactly as 
\begin{align}\label{channel_mat}
\bfH_s=\sum_{l = 1}^L \rho_{s,l} \bfa(\theta_l) \bfb(\tau_l)^\herm.
\end{align}
The rows of $\bfH_s$ correspond to the antenna elements, whereas its columns correspond to the OFDM subcarriers. 
The vectorized channel vector $\bbh_s=\vec(\bfH_s)$ is, therefore, given by
\begin{align}\label{channel_mat_vec}
\bbh_s =\sum_{l=1}^L \rho_{s,l} \bba(\theta_l,\tau_l),
\end{align}
where $\bba(\theta,\tau) = \vec(\bfa(\theta)\bfb(\tau)^\herm) \in \bC^{MN}$ denotes the array response in the angle-delay $(\theta,\tau)$.
Since the MPC coefficients $\rho_{s,l} \sim \cg(0, \sigma_{l}^2)$ are independent and circularly symmetric Gaussian variables, from \eqref{ziocanale2} it is 
immediate to check that the statistics of $\bfh_s[\omega]$ are invariant under circular shifts (with period $N$) in $\omega$,
implying stationarity in the frequency domain. 


\subsection{Antenna-Frequency Sampling}\label{chmod_subsamp}

As explained in Section \ref{sec:pilot-contamination}, without any loss of generality, we can assume that a UL pilot sequence for each user probes its channel over a subset of subcarriers in an individual OFDM symbol. Also, the pilot corresponding to different users are sent either across different OFDM symbols (disjoint in time) or across the same OFDM symbol but on disjoint set of subcarriers (disjoint in frequency). 
As before, we focus on a single user and denote by $\clI^\ttf_s=\{c^\ttf_{s,1}, \dots,c^\ttf_{s,n}\}$ the indices of the subcarriers acquired for this user at slot $s$. In addition, we consider the general case where also the antennas may be subsampled. This is done for the sake of generality, and also because 
one may wish to exploit the channel spatial correlation and reduce the sampling overhead at the receiver side. We denote by $\clI^\tta_s=\{c^\tta_{s,1}, \dots, c^\tta_{s,m}\}$ 
the indices of the antennas sampled at time slot $s$. We define $n\times N$ and $m \times M$
selection (or sampling) matrices\footnote{{In this paper, for simplicity, we focus on 0-1 antenna and subcarrier sampling matrices. This type of sampling is suitable for the Compressed Sensing algorithm that we develop later on in the paper to estimate the angle-delay PSF. However, our proposed method can be extended to work with more general projection matrices in the antenna and also subcarrier domain.}} 
 ${\bfS_s^\tta}$ and ${\bfS_s^\ttf}$, 
where $[{\bfS_s^\tta}]_{i, c^\tta_{s,i}}=1$ and $[{\bfS_s^\ttf}]_{j, c^\ttf_{s,j}}=1$, for $i\in [n]$ and $j \in [m]$. The sampled channel matrix at slot $s$ is given by 
$\widetilde{\bfH}_s={\bfS_s^\tta} \bfH_s {{\bfS_s^\ttf}}^\herm$. Using the $\vec$ notation, this can be written as 
\begin{align}\label{eq_vec_samp}
\widetilde{\bbh}_s & = \bS_s  \bbh_s, 
\end{align}
where $\bS_s={\bfS_s^\ttf} \otimes {\bfS_s^\tta}$ is of dimension $mn \times MN$, and where 
we used the well-known identity $\vec(\bfA \bfB \bfC)=(\bfC^\transp \otimes \bfA) \vec(\bfB)$. 
Notice that $\bS_s \bS_s^\herm =\bfI_{mn}$, and 
that $\bS_s$ has only a single element equal to $1$ in each row at column indices given by
\begin{align}\label{I_set}
\clI_s:=\{M(c_s^\ttf -1)+c_s^\tta: \ \ c_s^\ttf \in \clI_s^\ttf,\ c_s^\tta \in \clI_s^\tta\} \subseteq [MN].
\end{align}
{Using the above notation, the observation at the reference BS corresponding to a generic user (see (\ref{UL-pilot-measurement}) and \eqref{intro_wideband_sketch}) can be 
written as  $\bbx_s=\bS_s \bby_s$, where 
\begin{equation} \label{UL-pilot-measurement-vec}
\bby_s =  \bbh_s+ \sum_{j\in \clK} \bbh_{j,s} + \bbz_s,
\end{equation}
where $\Kc$ denotes the set of \cop s of a generic user in the reference cell/sector, and where, for notation simplicity, we dropped the index of the user and the copilot set $\clK$ (see, e.g., \eqref{pilot_cont_new} and \eqref{UL-pilot-measurement}) and 
 indicated the channel vectors of a generic user and of its \cop s at slot $s$ by $\bbh_s$ and $\bbh_{j,s}$, $j \in \clK$.  
}

\section{Estimation of Sparse Scattering Channel}\label{sec:ch_estim}

In this section, we propose a low-complexity algorithm to estimate the sparse geometry of the channel in the angle-delay domain as illustrated in Fig.\,\ref{multicell}. The resulting estimator is used in  Section \ref{sec:PDC} to perform pilot decontamination and channel interpolation.

\subsection{Low-dim Signal Structure}

Consider the reference user-BS pair with channel 
at slot $s$ given by \eqref{channel_mat_vec}. The covariance matrix of $\bbh_s$ is given by $\bfC_{\bbh}:=\bE[\bbh_s \bbh_s^\herm]=\sum_{l=1}^L \sigma_{l}^2 \bba(\theta_l,\tau_l) \bba(\theta_l,\tau_l)^\herm$.  
It is seen that although $\bfC_{\bbh}$ is a very large-dim $MN\times MN$ matrix, it is very low-rank (here the rank is $L$), 
due to sparse angle-delay scattering. This low-rank property still holds when the channel consists of a continuum of MPCs, provided 
that they have a small angle-delay support. 

In the UL pilot observation model in (\ref{UL-pilot-measurement-vec}), we denote by 
$
\bbd_s=\bbh_s+ \sum_{j\in \clK} \bbh_{j,s}
$
the \textit{superposition of the channel vectors} (SCVs) of the desired user and that of its $|\Kc|$  \cop s. 
Because of the distance-dependent pathloss, the number of \cop s  with a significant received power is quite small. 
In particular, all the \cop s with covariance matrices $\bfC_{\bbh_j}:=\bE[\bbh_{j,s} \bbh_{j,s}^\herm]$, $j\in \clK$, for which $\frac{1}{MN} \trace(\bfC_{\bbh_{j}}) \ll \sigma^2$ can be neglected.
Hence, without loss of generality, we can restrict $\Kc$ to include only the \cop s with significant ``raise over thermal'', i.e., those whose  
received power at the reference BS is significantly larger than the noise level. 
Therefore,  the covariance matrix $\bfC_{\bbd} := \bE[\bbd_s \bbd_s^\herm] = \bfC_{\bbh} + \sum_{j\in \clK} \bfC_{\bbh_{j}}$ of SCVs
is still very low-rank. 
 
Our goal in this section is to exploit this low-rank structure to estimate $\bfC_{\bbd}$ efficiently. 
To do so,  we collect multiple sketches  $\bbx_s=\bS_s \bby_s$, via $mn\times MN$ possibly time-variant sampling operators 
$\bS_s$, inside a window of size $\sfw$ of training slots across $\sfw$ CBs. 
We  represent these sketches by an $mn\times \sfw$ matrix $\bX$. Recall that the sampling matrix $\bS_s={\bfS_s^\ttf}\otimes {\bfS_s^\tta}$  consists of antenna and frequency sampling, where $\bfS^{\ttf}_s$ samples some of the subcarriers of a pilot OFDM symbol according to the UL pilot pattern (0-1 pattern) assigned to the user (see Section \ref{sec:pilot-contamination}), and where 
${\bfS_s^\tta}$ samples some of the antennas (pseudo)-randomly in each slot $s$. 
The performance of our proposed subspace estimation algorithm improves if the frequency signature of the user is also 
non-equally spaced and (pseudo)-randomly time-varying over the slots.
This can be implemented in practice by assigning a frequency-hopping pseudo-random pilot pattern to the users synchronized with the BS, 
analogous to what is currently done in CDMA systems. 
The drawback is that, in contrast with the uniform sampling scheme suggested by the classical 
Shannon-Nyquist sampling, the recovery of the whole instantaneous channel matrix from its nonuniform samples requires 
more complicated interpolation algorithms. As we will explain in Section \ref{PDC_opt} and \ref{sec:masking}, our proposed channel interpolation 
technique can be easily applied to both uniform and nonuniform sampling cases without incurring any additional complexity for the nonuniform one. 
The design of suitable pseudo-random frequency signatures yielding  easy interpolation is itself an interesting problem, which is beyond the scope of this paper. 

\subsection{Low-Complexity Subspace Estimation}\label{convex_equi}

We use the low-complexity algorithm we developed in our previous work \cite{haghighatshoar2017massive, haghighatshoar2016low} to estimate the signal subspace of the SCVs $\bbd_s$ from the sketches $\{\bbx_s: s\in[\sfw]\}$. 
The proposed algorithm is reminiscent of \textit{Multiple Measurement Vectors} problem in Compressed Sensing and exploits the joint sparsity of SCVs 
in the angle-delay domain.  We first quantize the angle-delay domain into a discrete grid $\clG:=\{(\theta_i, \tau_i)\}$, where for simplicity we use a uniform rectangular grid with $G=G^\theta G^\tau$ elements, with corresponding oversampling factors $\frac{G^\theta}{M}$ and $\frac{G^\tau}{N}$ in the angle and the delay domains, respectively.
We define an $MN\times G$ quantized dictionary matrix $\bA$ whose $i$-th column is given by $\overline{\bba}(\theta_i, \tau_i)$, where $\overline{\bba}(\theta, \tau)=\frac{\bba(\theta, \tau)}{\sqrt{MN}}$ denotes the normalized array response at angle-delay $(\theta,\tau)$. 
We define the $mn\times \sfw$ matrix $\bX=[\bbx_1, \dots, \bbx_\sfw]$ that contains the sketches $\{\bbx_s:s \in[\sfw]\}$. We assume that the noise power $\sigma^2$ in each antenna is known and normalize the received sketches by $\sigma$ where, 
for simplicity of notation, we denote the normalized sketches $\frac{\bX}{\sigma}$ again by $\bX$. 
We use the following $l_{2,1}$-norm regularized least squares proposed in \cite{haghighatshoar2016low} to estimate the signal subspace of the channel superposition:
\begin{align}\label{eq:l2_1_second}
\bfW^*=\argmin_{\bfW}\frac{1}{2}\sum_{s=1}^{\sfw} \|\widetilde{\bA}_s \bfW_{.,s} - \bX_{.,s}\|^2 +  \sqrt{\sfw}  \|\bfW\|_{2,1},
\end{align}
where $\widetilde{\bA}_s =\sqrt{\frac{MN}{mn}} \bS_s  \bA$ is a scaled and subsampled (via $\bS_s$) version of $\bA$, and where $\bfW \in \bC^{G\times \sfw}$ is a matrix whose rows correspond 
to the random channel gain of the MPCs over the quantized grid $\clG$ across $\sfw$ slots. Notice that 
$\|\bfW\|_{2,1}=\sum_{i=1}^G \|\bfW_{i,.}\|$ denotes the $l_{2,1}$-norm of $\bfW$ with $\bfW_{i,.} \in \bC^\sfw$ denoting the $i$-th 
row of $\bfW$.  The sparsity of the SCVs in the angle-delay domain results in the row-sparsity of the coefficient 
matrix $\bfW$, i.e., $\bfW$ must have only a few nonzero rows along the active grid elements $(\theta_i, \tau_i)\in \clG$ corresponding
 to the MPCs. 

In our previous work \cite{haghighatshoar2016low}, we used a $l_{2,1}$-norm regularizer for $\bfW$ to promote this row-sparsity. The resulting algorithm
is recalled here since it forms a key step of the proposed channel decontamination and interpolation scheme. 
\begin{algorithm}[t]
\caption{Forward-Backward Splitting with Nestrov's Update.}
\label{fb_alg_nest}
\begin{algorithmic}[1]
\State {{\bf Initialization:}} Fix $\bfW^{(0)}$, set $\bfZ^{(0)}=\bfW^{(0)}$, and $t_0=1$.
\For{$k=0,1,\dots,$}
\State $\bfR^{(k)}=\bfZ^{(k)} - \frac{1}{\beta} \nabla f_1(\bfZ^{(k)})$ and $\bfW^{(k+1)}=\prox_{\frac{1}{\beta} f_2} (\bfR^{(k)})$.
\State $t_{k+1}=\frac{1+\sqrt{4 t_k^2+1}}{2}$ and $\mu_k=1+\frac{t_k -1}{t_{k+1}}$.
\State $\bfZ^{(k+1)}=\bfW^{(k)} + \mu_k(\bfW^{(k+1)} - \bfW^{(k)})$.
\EndFor
\end{algorithmic}
\end{algorithm}
Consider the objective function \eqref{eq:l2_1_second}. After suitable scaling, we can write \eqref{eq:l2_1_second} as the minimization of function $f(\bfW)=f_1(\bfW)+f_2(\bfW)$, where $f_1(\bfW)=\frac{1}{2\zeta} \sum_{c=1}^\sfw \|\widetilde{\bA}_c \bfW_{.,c} - \bX_{.,c}\|^2$ with $\zeta=\sqrt{\sfw}$  and $f_2(\bfW)= \|\bfW\|_{2,1}$. 
The gradient of $f_1$ is a $G\times \sfw$ matrix $\nabla f_1(\bfW)$ whose $c$-th column, $c\in[\sfw]$, is 
given by $\nabla f_1(\bfW)_{.,c}=\frac{1}{\zeta} \widetilde{\bA}_c^\herm (\widetilde{\bA}_c \bfW_{.,c} - \bX_{.,c})$. 
To apply the algorithm in \cite{haghighatshoar2016low}, we need to compute the Lipschitz constant of $\nabla f_1$, i.e., the smallest constant $\beta>0$ such that for  every $\bfW, \bfW' \in \bC^{G\times \sfw}$:
\begin{align}
\|\nabla f_1(\bfW) - \nabla f_1(\bfW')\| \leq \beta \|\bfW - \bfW'\|.
\end{align}
We can check that $\beta\leq \frac{1}{\zeta}\max_{c \in [\sfw]} \lambda_{\max} \big \{ {\widetilde{\bA}_c^\herm \widetilde{\bA}_c} \big \}$, where $\lambda_{\max}$ denotes the maximum singular value of a given matrix. Note that if the grid size $G$ is sufficiently large and the grid points $\{(\theta_i, \tau_i)\}$ are distributed quite uniformly and densely over the angle-delay domain, we have that 
\begin{align*}
\widetilde{\bA}_c^\herm \widetilde{\bA}_c&=\frac{1}{mn} \bS_c \Big \{ \sum_{i=1}^G \bba(\theta_i, \tau_i) \bba(\theta_i, \tau_i) ^\herm\Big \}\bS_c^\herm\approx \frac{G}{mn} \bS_c \bfI_{MN} \bS_c^\herm= \frac{G}{mn} \bfI_{mn},
\end{align*}
where we used $\bS_c \bS_c^\herm=\bfI_{mn}$. This implies that $\beta\approx \frac{G}{\zeta mn}=\frac{G}{ mn \sqrt{\sfw}}$. We also need the proximal operator of $l_{2,1}$-norm $f_2$ with a scaling $\alpha>0$ defined by $\prox_{\alpha f_2}: \bC^{G\times \sfw} \to \bC^{G\times \sfw}$, whose $i$-th row is given by
\begin{align}
(\prox_{\alpha f_2}(\bfW))_{i,.}= \frac{(\|\bfW_{i,.}\|-\alpha)_+}{\|\bfW_{i,.}\|} \bfW_{i,.}
\end{align}
and corresponds to a \textit{shrinkage} operator shrinking the rows of $\bfW$ by $\alpha$, where  $(x)_+ \eqdef \max(x,0)$. The algorithm proposed in \cite{haghighatshoar2016low} with the Nestrov's step-size update is given by Algorithm \ref{fb_alg_nest}.
We have also the following performance guarantee from \cite{haghighatshoar2016low}.
\begin{proposition}[{\cite[Theorem 11.3.1]{nemirovski2005efficient}}]\label{prop:nest}
Let $\{\bfW^{(k)}\}_{k=0}^\infty$ be the sequence generated by Algorithm \ref{fb_alg_nest} for an arbitrary initial point $\bfW^{(0)}$ and for the step-sizes according to the Nestrov's update rule. Then, for any $k$, we have $f(\bfW^{(k+1)}) - f(\bfW^{*}) \leq \frac{4 \beta \|\bfW^*-\bfW^{(0)}\|^2}{(k+1)^2}$.
\end{proposition}
Let $\bfW^*$ be the optimal solution of \eqref{eq:l2_1_second} and let $s_i=\frac{1}{mn} \|\bfW^*_{i,.}\|$ be the $l_2$-norm of the $i$-th 
row of $\bfW^*$.  The covariance matrix $\bfC_{\bbd}$ of the channel superposition can be estimated from \cite[Proposition 1]{haghighatshoar2016low} by 
\begin{align}\label{C_approx}
\bfC^*_{\bbd}\approx \sum_{i=1}^G s_i \overline{\bba}(\theta_i,\tau_i)\overline{\bba}(\theta_i,\tau_i)^\herm.
\end{align}

\begin{remark}\label{rem:slow}
By increasing the grid size $G$, \eqref{C_approx} provides a more precise estimate of the covariance matrix of SCVs. However, as also mentioned in \cite{haghighatshoar2016low}, since the Lipschitz constant $\beta=\frac{G}{ m n\sqrt{\sfw}}$ grows proportionally to $G$, it is seen from Proposition \ref{prop:nest} that increasing $G$ reduces the convergence speed of  the algorithm. Intuitively, by increasing $G$ and as a result $\beta$, the shrinkage operator $\prox_{\frac{1}{\beta} f_2}$ in Algorithm \ref{fb_alg_nest} becomes softer, and the algorithm requires more iterations to converge, although at the end the resulting estimate $\bfC^*_{\bbd}$ in \eqref{C_approx} is generally improved. 
\end{remark}

\subsection{Computational Complexity}\label{sec:comp_complexity}

Each iteration of Algorithm \ref{fb_alg_nest} requires computing $\sfw$ columns of $G\times \sfw$ gradient $\nabla f_1$, where the $c$-th column, $c \in [\sfw]$, is given by $\nabla f_1(\bfW)_{.,c}=\frac{1}{\zeta}\widetilde{\bA}_c^\herm (\widetilde{\bA}_c \bfW_{.,c} - \bX_{.,c})$, evaluated at $\bfW=\bfW^{(k)}$ at  iteration $k$. 
Here, we consider a special grid $\clG$ whose discrete AoAs $\theta_k$ belong to
\begin{align*}
\Theta:=\big \{\sin^{-1}\big((-1+\frac{2(i-1)}{G}) \sin(\theta_{\max})\big): i\in [G^\theta]\big \},
\end{align*}
in the angular range $[-\theta_{\max}, \theta_{\max}]$. We also assume that all the discrete delay elements $\tau_k$ in $\clG$ belong to a uniform grid in the delay domain $[0, \Delta \tau_{\max}]$ of size $G^\tau$. For this particular choice of the grid $\clG$, the gradient matrix $\nabla f_1(\bfW)$ can be efficiently 
computed via 2D \textit{Fast Fourier Transform} (FFT) as follows. 
Let $\clI_{c}^{\tta} \subseteq [M]$ and $\clI_{c}^{\ttf} \subseteq [N]$ denote the indices of the sampled antennas and subcarriers in the OFDM symbol at $c \in [\sfw]$. For each $c\in [\sfw]$, we first compute $\widetilde{\bA}_c \bfW_{.,c}$. Following the \matlab notation, we first set  $\bfM=\mathsf{reshape}(\bfW_{.,c}, G^\theta, G^\tau)$, and let $\widetilde{\bfM}=G^\theta G^\tau\, \mathsf{ifft2}(\bfM)$ be the inverse 2D \textit{Discrete Fourier Transform} (DFT) of $\bfW_{.,c}$ scaled with $G$. This can be efficiently computed using the FFT algorithm in $O(G \log_2(G))$ operations under mild conditions on the integers $G^\theta$ and $G^\tau$ (e.g., they may be 
powers of $2$).  Then, $\widetilde{\bA}_c \bfW_{.,c}$ is simply given by $\frac{1}{\sqrt{mn}} \vec(\widetilde{\bfM}(\clI_{c}^{\tta}, \clI_{c}^{\ttf}))$. The whole complexity of this step is $O\big(\sfw G \log_2(G) \big )$. 
After computing $\widetilde{\bA}_c \bfW_{.,c}$, we need to calculate $\widetilde{\bA}_c^\herm \bfR_{.,c}$, where $\bfR$ is an $mn\times \sfw$ matrix with $\bfR_{.,c}=\widetilde{\bA}_c \bfW_{.,c} - \bfX_{.,c}$, for $c\in[\sfw]$. To do this, we set $\bfM$ to be an $M\times N$ all-zero matrix and embed $\bfR_{.,c}$ in $\bfM$ in indices belonging to $\clI_{c}^{\tta}$ and $\clI_{c}^{\ttf}$ such that $\vec \big( \bfM(\clI_{c}^{\tta}, \clI_{c}^{\ttf}) \big )=\bfR_{.,c}$, and take the 2D DFT of $\bfM$, which gives $\widetilde{\bA}_c^\herm \bfR_{.,c}=\frac{1}{\sqrt{mn}} \vec \big ( \mathsf{fft2}(\bfM, G^\theta, G^\tau) \big )$. The whole complexity of this step is again $O\big(\sfw G \log_2(G) \big )$.

Letting $T_\mathsf{conv}$ be the number of iterations necessary for the convergence, the whole computational complexity of our algorithm is $O\big(2 T_\mathsf{conv} \sfw G \log_2(G) \big )$. Typically, $T_\mathsf{conv}$ scales proportionally to $\frac{G}{MN}$ where, as also explained in Remark \ref{rem:slow}, increasing the grid size $G$ slows down the convergence of the algorithm. 
We always use $\frac{G^\theta}{M}=\frac{G^\tau}{N}=2$. Our numerical simulations show that for this choice of the oversampling factor, Algorithm \ref{fb_alg_nest} runs quite fast and converges in only a few iterations even for quite large $M,N\approx 256$ and $\sfw \approx 200$. 

{
\subsection{System-level Considerations}
As we will further explain in the simulations in Section \ref{sec:sim}, our 
proposed algorithm  is able to extract the signal subspace of SCVs 
by gathering UL pilot observations over a  time window of the order of $50$\,ms, 
over which the underlying signal subspace can be safely assumed to remain invariant. 
In almost all practical situations, the subspace remains stable for a time scale of the order of $1\sim 10$\,s 
(see Section \ref{sec:wss} and \cite{tse2005fundamentals,liu2012cost}), 
which is much larger than the time scale required for estimating the subspace. Thus, the estimated subspace can be used for many time slots. 
In addition, the estimation phase results in almost no system-level overhead since i)\,during the estimation phase the system can still work 
in the (standard) contaminated mode ii)\,the sketches of the wideband channel vectors of the users gathered for subspace estimation are also required 
for serving the users, so subspace estimation does not impose effectively any additional sampling or pilot transmission overhead. 
In practice, one can apply subspace tracking algorithms as in \cite{haghighatshoar2016low} to update the estimated subspace upon arrival 
of each new observation at each new time slot $s$. This has the additional advantage that the computational complexity of the subspace 
estimation is distributed across several time slots. 
}


\section{Pilot Decontamination and Channel Interpolation}\label{sec:PDC}

\subsection{Estimating the Angle-Delay Power Spread Function}\label{PDC_psd}

Let $\bfW^*$ be the $G\times \sfw$ matrix of coefficients obtained as the optimal solution of the optimization \eqref{eq:l2_1_second}. We define a discrete positive measure $\gamma(\theta, \tau)$ over the angle-delay domain that assigns the weight $\gamma(\theta_l, \tau_l)=\frac{1}{{mn}} \|\bfW^*_{l,.}\|$ to the $l$-th grid element $(\theta_l,\tau_l) \in \clG$. From Section \ref{convex_equi}, the covariance matrix $\bC_{\bbd}$ of  SCVs is well-approximated by $\bC^*_{\bbd}$ in \eqref{C_approx}, which can be written as $\bC^*_{\bbd}=\int \gamma(d\theta, d\tau) \bba(\theta,\tau) \bba(\theta,\tau)^\herm$ in terms of the discrete measure $\gamma(\theta, \tau)$. Consequently, we expect that $\gamma(\theta,\tau)$ be a good approximation of the \textit{angle-delay Power Spread Function} (PSF) of   SCVs. 
We also define the marginal measure $\gamma(\theta)=\int_0^{\Delta \tau_{\max}} \gamma(\theta, d\tau)$, which provides an approximation 
of the PSF in the angle domain $\theta$.

\subsection{Clustering in the Angle-Delay Domain}\label{sec:cluster}

\begin{figure}[h]
\centering
\includegraphics[width=3.5cm]{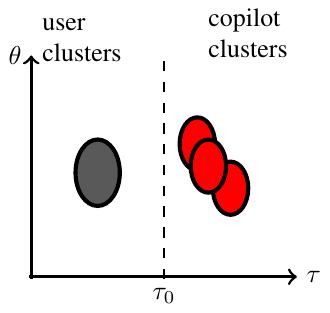}
\caption{Illustration of an unsupervised clustering using the a priori information in the delay domain.} 
\label{fig:clst_unsup}
\end{figure}

A crucial ingredient of our pilot-decontamination method is a \textit{clustering algorithm} in the angle-delay domain. 
The output of such a clustering algorithm is a decomposition 
$\big(\gammas(\theta, \tau), \gammai(\theta, \tau)\big )=\clst(\gamma(\theta,\tau))$ of the angle-delay PSF $\gamma(\theta,\tau)$ into a desired 
signal part $\gammas(\theta,\tau)$, corresponding to the wideband channel vector of the desired user, and a copilot 
interference part $\gammai(\theta,\tau)$, corresponding to the superposition of the wideband channel vectors of  \cop s. 
This can be done using supervised or unsupervised learning techniques.
In an unsupervised scheme, $\clst$  exploits only the a priori knowledge about the desired user and its \cop s. 
This is typically based on the geometric constraints of the cell in which the user signal propagates such as the location 
as well as the received power strengths of different delay-angle elements. For example, if it is a priori known that all the copilot 
MPCs are separable in the delay domain, say by a delay threshold $\tau_0\in [0, \Delta \tau_{\max})$, then the clustering algorithm $\clst$ can be as simple as $\gammas(\theta,\tau)= \gamma(\theta,\tau){\bf 1}_{\{\tau \in [0,\tau_{0}]\}}$, and $\gammai(\theta,\tau)= \gamma(\theta,\tau){\bf 1}_{\{\tau \in (\tau_{0},\Delta \tau_{\max}]\}}$, where ${\bf 1}_{\clB}$ denotes the indicator of a set $\clB$, as illustrated in  Fig.\,\ref{fig:clst_unsup}.

\begin{figure}[t]
\centering
\includegraphics[width = 9cm]{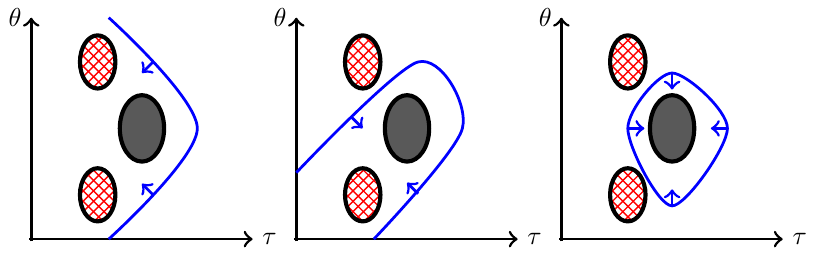}
\caption{ Illustration of a supervised clustering algorithm. In this figure, $\gamma(\theta,\tau)$ consists of one signal cluster (solid gray) plus two interference clusters (dashed) corresponding to two \cop s.}
\label{fig:clst_sup}
\end{figure}

In a supervised scheme, $\clst$ has access to an ``oracle'' provided by higher communication layers, 
which can be exploited to perform adaptive clustering.
In brief, $\clst$ starts from an initial clustering and refines it iteratively using the oracle response 
until a good partition of $\gamma(\theta,\tau)$ into $\gammas(\theta, \tau)$ and $\gammai(\theta, \tau)$ is obtained. 
A simple example of this is illustrated in Fig.\,\ref{fig:clst_sup}. In this example, $\gamma(\theta, \tau)$ consists of one signal cluster 
and two \cop\  clusters, where for simplicity we have assumed that these clusters are non-overlapping.  
The $\clst$ starts with the obvious initialization $\gammas(\theta,\tau) = \gamma(\theta, \tau) $, i.e., that there is no contamination (left figure in 
Fig.\,\ref{fig:clst_sup}).  Based on this assumption, it estimates the channel vector on all the subcarriers (using the channel interpolation scheme proposed in the following), and based on this channel estimation it attempts to decode the UL user data. In the presence of significant contamination, the effective \textit{Signal to Interference plus Noise Ratio} (SINR) is degraded and  some standard link layer control mechanism detects the data packet in error. 
This error detection mechanism can be exploited as an \textit{oracle} for supervised learning. 
In the presence of a packet error, the $\clst$ tries a different selection of the clusters (e.g., as in the center figure in Fig.\,\ref{fig:clst_sup}). 
The process is repeated until the data packet is decoded correctly. Notice that in this case, although there is no guarantee that {\em all} 
the copilot interference be removed,  we have the guarantee that it has been removed enough to decode the data, whenever this is
possible. This means that the effective SINR for the desired user is large enough to
achieve successful decoding. Of course, if no successful decoding is achieved after a fixed number of iterations, the packet is rejected, and the desired user is re-scheduled for transmission on a later slot. This is not different from a standard ``packet failure'' event, 
which is handled by retransmission or by any suitable upper layer protocol in a completely standard manner. 
Notice also that $\clst$ learns the suitable  
clustering without any explicit feedback from the users since the whole process is performed entirely at the BS receiver on a single packet detection. 
Therefore, it does not  involve any additional latency with respect to a standard massive MIMO system. 
Interestingly, in this example, the clusters corresponding to the \cop s have smaller propagation delays than the one corresponding to the desired user. 
As a result,  the previously mentioned unsupervised algorithm, which only exploits the propagation delay of the users, would fail to 
identify the signal cluster. Such a situation arises, for example, in a cell-free massive MIMO system, where copilot interference may be particularly harmful \cite{ngo2015cell, bursalioglu2016rrh}.


\subsection{Instantaneous Channel Estimation/Interpolation and Pilot Decontamination}\label{PDC_opt}
Let $\gammas(\theta,\tau)$ and $\gammai(\theta, \tau)$ be the PSFs of the desired user  and of the \cop s, obtained as described before. 
For channel decontamination and interpolation, we apply the MMSE smoothing filter (\ref{MMSE-smoothing}) with ``plug-in'' covariance estimates
given by $\bfC^*_{\bbh} = \int \gammas(d\theta,d\tau) \bba(\theta,\tau) \bba(\theta,\tau)^\herm$
for the desired user channel, and by $\bfC^*_{\bbi} = \int \gammai(d\theta,d\tau) \bba(\theta,\tau) \bba(\theta,\tau)^\herm$ 
for the superposition of the \cop\  channels.  
The resulting plug-in channel estimator-interpolator is given by 
\begin{align}\label{mmse_subsample}
\widehat{\bbh}_s &= \Sigmam_{\bbh \bbx} \bfC^{-1}_{\bbx} \bbx_s\approx \bfC^*_{\bbh} \bS_s^\herm (\sigma^2 \bfI_{mn} + \bS_s ( \bfC^*_{\bbh}   + \bfC^*_{\bbi} ) \bS_s^\herm)^{-1} \bbx_s,
\end{align}
where $\Sigmam_{\bbh \bbx}=\bE[\bbh_s \bbx_s^\herm]$ denotes the cross covariance matrix of $\bbh_s$ and $\bbx_s$, and where $\bbx_s = \vec( \Xm_s)$ is the UL pilot observation at time slot $s$.
Under the condition that the estimated covariance matrices 
$\bfC^*_\bbh$ and $\bfC^*_\bbi$ are close to the true covariance matrices $\bfC_\bbh$ and $\bfC_\bbi$, 
the channel estimator in (\ref{mmse_subsample}) is close to the ideal MMSE smoothing filter \eqref{MMSE-smoothing}. 
Notice that in the absence of copilot interference (i.e., for $\bfC_\bbi = {\bf 0}$)
such  MMSE smoothing filter implements the optimal channel interpolation in the antenna and frequency domain in the MMSE sense. 
In conventional implementations, ``ad-hoc'' channel interpolation techniques in the OFDM subcarrier domain are used in order to
interpolate the unobserved columns (subcarriers) and rows (antennas) 
of the channel matrix $\bfH_s$  from the instantaneous noisy UL pilot observation $\Xm_s$ as given in (\ref{UL-pilot-measurement}). 
Typical schemes include simple piecewise constant,  linear, or DFT-based (Sinc-shaped) interpolation (see \cite{choi2005optimum, hutter1999channel} 
and the refs. therein).  The advantage of  our proposed subspace estimation for pilot decontamination 
is that, as seen from \eqref{mmse_subsample}, the channel vector $\bbh_s$ can be directly estimated from the 
sketch $\bbx_s$,  thus, we obtain per-slot channel estimation/interpolation for free.

\section{Low-complexity Channel Interpolation and Pilot Decontamination}\label{PDC_subopt}

Computing the $MN\times MN$ covariance matrices from the estimated PSF $\gamma(\theta,\tau)$ and performing the matrix multiplication for the MMSE estimation in \eqref{mmse_subsample}, as proposed in the previous section, may result in a prohibitive complexity 
for typical massive MIMO systems (e.g., $M = 128$ antennas and $N = 128$ subcarriers).  In this section, we propose two low-complexity 
algorithms to address this computational complexity issue. 
The first algorithm, explained in Section \ref{sec:masking}, uses a masking technique in the angle-delay domain, which yields a low-complexity approximation of the MMSE 
estimator proposed in \eqref{mmse_subsample}. 
The second algorithm, stated in Section \ref{sec:angle_separ}, has much lower complexity but, in order to guarantee to eliminate pilot contamination, 
 requires a stronger angular separability condition as we will explain. 

\subsection{Interpolation and Pilot Decontamination by Masking}\label{sec:masking}

Let $\gamma(\theta,\tau)$ be the estimated PSF supported on the grid elements $(\theta_l,\tau_l) \in \clG$ as in Section \ref{PDC_psd}. We define the mask $\clM$ as follows 
\begin{align}\label{mask_M}
\clM:=\big \{(\theta_l,\tau_l): \gamma(\theta_l,\tau_l) \geq \iota \big\},
\end{align}
where $\iota \in \bR_+$ denotes a masking threshold in the angle-delay domain which selects only those grid elements with a significantly large received power. We assume that $\clM=\clM^\bbh\cup \clM^\bbi$ is decomposed into disjoint signal and \cop\ interference masks $\clM^\bbh$ and $\clM^\bbi$ with $\clM^\bbh \cap \clM^\bbi=\emptyset$  via the clustering algorithm $\clst$. For the case of supervised clustering, $\clst$ changes  
the masks $\clM^\bbh$ and $\clM^\bbi$ in each iteration, while keeping their union equal to $\clM$  as in \eqref{mask_M}, until it finds a good estimate 
of the true signal cluster, e.g., when the packet decoding is successful as described before.

Let ${\bfS_s^\tta}$ and ${\bfS_s^\ttf}$ be the antenna and frequency sampling matrices at slot $s$. We apply joint interpolation and pilot decontamination as follows. We find an estimate of the channel matrix $\bfH_s$ denoted by $\bfP$ and an estimate of copilot interference denoted by $\bfQ$ via minimizing $\|\bfX_s - {\bfS_s^\tta} (\bfP+\bfQ)  {{\bfS_s^\ttf}}^\herm\|$, where $\bfX_s={\bfS_s^\tta} \bfY_s {{\bfS_s^\ttf}}^\herm$ denotes the subsampled observations  at slot $s$ and where $\bfY_s$ denotes the noisy contaminated received wideband signal. To do so, we impose the additional constraint that a significant amount of power of $\bfP$ and $\bfQ$ be concentrated in the mask $\clM^\bbh$ and $\clM^\bbi$ in the angle-delay domain, respectively. 
We denote by $\fovs: \bC^{M\times N} \to \bC^{G^\theta\times G^\tau}$ the oversampled 2D DFT and by $\ford: \bC^{G^\theta\times G^\tau}\to \bC^{G^\theta\times G^\tau}$ the usual 2D DFT in dimension $G^\theta\times G^\tau$, where $\frac{G^\theta}{M}$ and $\frac{G^\tau}{N}$ denote the oversampling factors in the angle and delay domain respectively. 
Note that for an $M\times N$ matrix $\bfH$, we have $\fovs(\bfH)=\ford(\bfH^\text{ovs})$ where $\bfH^\text{ovs}$ denotes a $G^\theta\times G^\tau$ matrix that has $\bfH$ in  its up-left corner and is zero elsewhere. This follows from the well-known property of DFT, where  
an oversampling in one domain can be obtained by zero-padding in the corresponding transform domain. 
For simplicity, we assume that $\ford$ is normalized such that it is an isometry preserving the matrix Frobenius norm, i.e., $\|\ford(\bfL)\|=\|\bfL\|$ for any $G^\theta\times G^\tau$ matrix  $\bfL$. We define the following  cost function for $\bfP$ and $\bfQ$
\begin{align}\label{c_PQ}
c(\bfP,\bfQ)&=\frac{1}{2} \|\bfX_s - {\bfS_s^\tta} (\bfP+\bfQ) {{\bfS_s^\ttf}}^\herm\|^2 + \ind_{\clM^\bbh}(\fovs(\bfP))+ \ind_{\clM^\bbi}(\fovs(\bfQ))
\end{align} 
where $\ind_{\clM^\bbh}, \ind_{\clM^\bbi}: \bC^{G^\theta\times G^\tau} \to \bR_+ \cup \{+\infty\}$ are convex regularizers penalizing those nonzero coefficients of their arguments not belonging to the masks $\clM^\bbh$ and $\clM^\bbi$ respectively. A simple regularizer is the indicator function of a mask $\clD$, given by:
\begin{align}\label{ind_def}
\ind_{\clD}(\bfK):=\left \{ \begin{array}{ll} \infty & \text{ if } \bfK({\sim\clD})\neq {\bf 0},\\ 0 & \text{otherwise,} \end{array} \right.
\end{align}
where $\bfK({\sim\clD})$ denotes those elements of the matrix $\bfK$ not belonging to $\clD$. The cost function in \eqref{c_PQ} is convex and its globally optimal solution $(\bfP^*, \bfQ^*)$ can be found via convex optimization techniques. The optimal solution $\bfP^*$ of \eqref{c_PQ} is an estimate of the decontaminated channel matrix $\bfH_s$. In the presence of antenna and frequency sampling, this technique (masking and optimization) provides an interpolation scheme to recover the whole channel matrix from its subsamples. 
In Appendix \ref{sec:admm}, we propose a low-complexity algorithm for solving \eqref{c_PQ} using \textit{Alternating Direction Method of Multipliers} (ADMM), which estimates/interpolates the decontaminated channel matrix with a complexity $O(G \log_2(G))$, where $G=G^\theta G^\tau$ denotes the total number of points in the grid $\clG$. This provides a low-complexity implementation of the   MMSE smoothing filter proposed in Section \ref{PDC_opt}.

\subsection{Low-complexity Pilot Decontamination under the Angular Separability Condition}\label{sec:angle_separ}
\begin{figure}[h]
\centering
\subfloat[Overlapping.\label{angular_overlapping}]{%
\includegraphics[width=3.5cm]{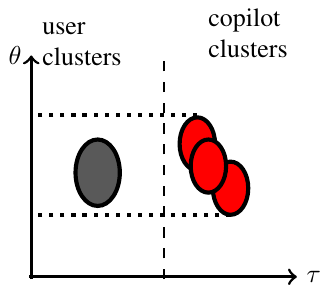}
}
\hspace{2cm}
\subfloat[Non-overlapping.\label{angular_nonoverlapping}]{%
\includegraphics[width=3.5cm]{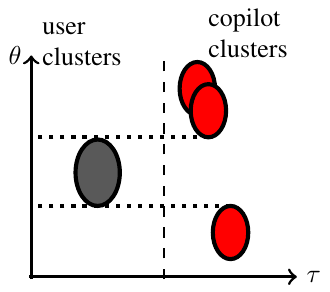}
}
\caption{Illustration of two extremes of angular overlap of the PSF of the desired 
user $\gammas(\theta)$ and that of its \cop s $\gammai(\theta)$: Overlapping (a) and Non-overlapping (b).}
\label{fig:PD_angular}
\end{figure}

In this section, we explain another pilot decontamination algorithm that has much lower complexity than the MMSE estimator \eqref{mmse_subsample} proposed in Section \ref{PDC_opt} but, to eliminate pilot contamination, it requires a stronger condition that marginal PSF $\gammas(\theta)$ and $\gammai(\theta)$ of the user and 
its \cop s have approximately disjoint supports in the angular domain, where we define $\gammas(\theta)=\int_0^{\Delta \tau_{\max}} \gammas(\theta,d\tau)$ with a similar definition holding for $\gammai(\theta)$. This is illustrated qualitatively in  Fig.\,\ref{fig:PD_angular}. 
%
%
%
Notice that the separability of $\gammas(\theta,\tau)$ and $\gammai(\theta,\tau)$ in the joint angle-delay domain is still necessary to successfully decompose (cluster) the PSF $\gamma(\theta, \tau)$ into its signal and interference components $\gammas(\theta,\tau)$ and $\gammai(\theta,\tau)$. 

Let $\bfH_s$ and $\bfH_{s,j}, j \in \Kc$, be the channel matrices of a user and of its \cop s, and let $\interF_s=\sum_{j \in \Kc} \bfH_{s,j}$ be the channel matrix of \cop\ interference. Let  $\bfC_{\bbh}^*=\int \gammas(d\theta d\tau) \bba(\theta,\tau) \bba(\theta,\tau)^\herm$ be the estimated covariance matrix of the channel vector $\bbh_s=\vec(\bfH_s)$ from the estimated PSF $\gammas(\theta,\tau)$ obtained from the clustering. It is not difficult to check that $\bfC^*_{\bbh}$  is a block-Toeplitz matrix, which implies that every column $\bfh_s[\omega]$ 
of $\bfH_s$ is an  $M$-dim Gaussian vector with a covariance matrix well approximated by $\bfC^*_{\bfh}= \int \gammas(d\theta) \bfa(\theta) \bfa(\theta)^\herm$, where $\bfC^*_{\bfh}$ is an $M\times M$ Toeplitz matrix and corresponds to the diagonal block of $\bfC^*_{\bbh}$. 
Similarly, every column $\interf_s[\omega]$ of the \cop \, 
interference $\interF_s$ is an $M$-dim Gaussian vectors with a Toeplitz covariance matrix given by $\bfC^*_{\interf}=\int \gammai(d \theta) \bfa(\theta) \bfa(\theta)^\herm$. 
Let $\bfY_s=\bfH_s + \interF_s + \bfZ_s$ be the received noisy  and pilot contaminated signal. For simplicity, we first assume that there is no antenna or frequency sampling and $\bfY_s$ is fully available. We consider the following suboptimal scheme for pilot decontamination: Instead of estimating the whole channel matrix $\bfH_s$ from $\bfY_s$,  as we did for the MMSE estimation in Section \ref{PDC_opt}, we estimate each column $\bfh_s[\omega]$ of $\bfH_s$ from the corresponding column $\bfy_s[\omega]$ of $\bfY_s$ individually. Since $\bfy_s[\omega]=\bfh_s[\omega]+\interf_s[\omega]+\bfz_s[\omega]$, this is a standard problem of estimating a Gaussian $M$-dim vector  $\bfh_s[\omega]$ in an additive colored Gaussian noise $\interf_s[\omega]+ \bfz_s[\omega]$. The resulting MMSE estimator can be simply written as 
\begin{align}\label{column_mmse}
\widehat{\bfh}_s[\omega]&=\Sigmam_{\bfh \bfy} \bfC_{\bfy}^{-1} \bfy_s[\omega]
\approx \bfC_{\bfh}^* \left (  \sigma^2 \bfI_M + \bfC_\bfh^* + \bfC_\interf^* \right )^{-1} \bfy_s[\omega],
\end{align}
where $\Sigmam_{\bfh \bfy}=\bE[\bfh_s[\omega] \bfy_s[\omega]^\herm]$ denotes the cross correlation matrix of $\bfh_s[\omega]$ and $\bfy_s[\omega]$. 
It is seen that the MMSE estimator  is an $M\times M$ linear operator, which requires computing the inverse of an $M\times M$ Toeplitz matrix rather than an $MN\times MN$ block-Toeplitz matrix, as was necessary for the joint MMSE estimator in Section \ref{PDC_opt}. More importantly, 
since the spatial correlation of the channel is invariant with the subcarrier index $\omega$ 
due to the stationarity in the frequency domain, the linear estimator is the same for all the columns 
of the channel matrix, thus, it needs to be computed only once. 

If in addition there is an antenna sampling via an operator $\bfS_s^\tta$, letting $\bfx_s[\omega]=\bfS_s^\tta \bfy_s[\omega]$ to be 
the $m$-dim sketch at subcarrier $\omega$ after antenna sampling, the MMSE estimator of $\bfh_s[\omega]$ from $\bfx_s[\omega]$ 
takes on the form 
\begin{align}\label{column_mmse_2}
\widehat{\bfh}_s[\omega]&=\Sigmam_{\bfh \bfx} \bfC_{\bfx}^{-1} \bfx_s[\omega]
\approx \bfC_{\bfh}^* {\bfS_s^\tta}^\herm \left  ( \sigma^2 \bfI_m + \bfS_s^\tta (\bfC_\bfh^* + \bfC_\interf^*) {\bfS_s^\tta}^\herm  \right )^{-1} \bfx_s[\omega].
\end{align}

When the channel matrices of several users are learned over the same OFDM symbol, only a subset of columns of $\bfY_s$ is observed for each user. In such a case, we apply the column-wise pilot-decontamination in \eqref{column_mmse} or \eqref{column_mmse_2} to estimate the corresponding columns of $\bfH_s$. Then, we apply traditional  channel interpolation methods to reconstruct the remaining columns of $\bfH_s$ from the estimated ones (e.g., via piecewise constant, linear, or DFT-based interpolation techniques \cite{choi2005optimum, hutter1999channel}).
The proposed suboptimal pilot decontamination reduces the implementation complexity considerably. However, the drawback is that in contrast with $\gammas(\theta,\tau)$ and $\gammai(\theta,\tau)$, which are usually well-separable in the joint angle-delay domain, $\gammas(\theta)$ and $\gammai(\theta)$ might generally overlap in the angle domain. In such a case, the dominant subspaces of $\bfC_{\bfh}$ and $\bfC_{\interf}$ will be highly overlapping, and the suboptimal MMSE will eliminate a significant fraction of the power of the columns of the channel matrix $\bfH_s$ lying in the interference subspace $\bfC_{\interf}$, which results in a poor design of the final beamforming matrix. 

\begin{figure}[t]
\centering
\subfloat[Before Shuffling the Pilots. \label{pilot_shuffling_overlap}]{%
\includegraphics[width=0.32\textwidth]{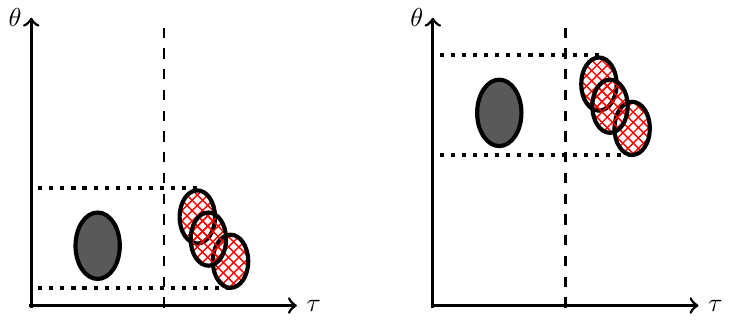}
}
\hspace{2cm}
\subfloat[After Shuffling the Pilots.\label{pilot_shuffling_nonoverlap}]{%
\includegraphics[width=0.32\textwidth]{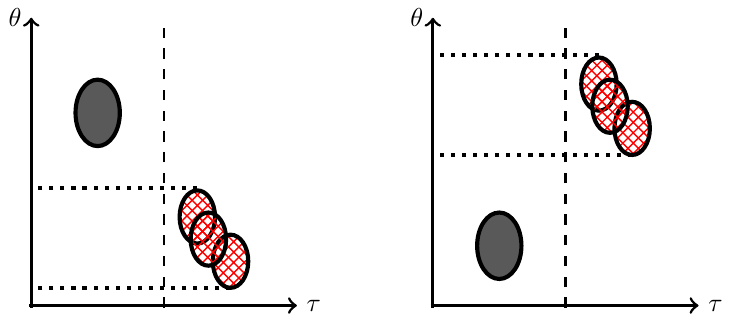}
}
\caption{Separability in the angle domain via shuffling the pilots among the users: Before shuffling the pilots (a) and after shuffling the pilots (b).}
\label{fig:pilot_shuffling}
\end{figure}

In practice, since the number of \cop s is typically small, if the users have a limited angular support and are quite randomly distributed inside the cell, there is a high chance that the effective overlap between $\gamma_s(\theta)$ and $\gamma_i(\theta)$ be quite negligible for most users. 
Another way to make $\gamma_s(\theta)$ and $\gamma_i(\theta)$ non-overlapping consists of shuffling 
the pilots assigned to the active users across the whole system as proposed in \cite{yin2013coordinated}. 
This is illustrated qualitatively  in Fig.\,\ref{fig:pilot_shuffling}, where by re-allocating the pilot of the users of interest the BS 
can induce angular separation with respect to the \cop s. 
Pilot-shuffling requires some coordination among neighboring BSs inside the system. In \cite{yin2013coordinated, bjornson2016pilot}, it is assumed that the  PSFs or the covariance matrices $\bfC_\bfh$ and $\bfC_\interf$ of all the users and their \cop s are available. In contrast, in this paper we estimate the PSFs by using wideband pilots and exploiting the sparsity in the angle-delay domain, and identify the signal and interference PSFs by applying suitable clustering algorithms. Hence, our scheme to identify $\bfC_\bfh$ and $\bfC_\interf$ directly 
from the pilot data  can be seen as an enabler for the coordinated pilot shuffling scheme in  \cite{yin2013coordinated} 
and the pilot decontamination in \cite{bjornson2016pilot}. 

\section{Simulation Results}\label{sec:sim}
In this section, we assess the performance of our proposed pilot decontamination and channel interpolation algorithm  via numerical simulations. 

\subsection{Cellular Geometry and Antenna Model}\label{power_model}
We consider a cellular system consisting of hexagonal cells of radius $R_\text{cell}=1.5$\,Km and a maximum tolerable delay spread of $\Delta \tau_{\max} =\frac{2R_\text{cell}}{c_0}=10$\,$\mu$s. 
For simulations, we assume the transmit/receive power decays with a power-loss exponent $\eta=3.2$ (for large cells), where the SNR before beamforming for a user located at a distance $r$ from the BS is given by $\text{SNR}(r)=\frac{\text{SNR}_{\max}}{1+(\frac{r}{r_0})^\eta}$, where $r_0=500$\,m, 
and where $\text{SNR}_{\max}$ is selected such that the SNR before beamforming for a user located at the cell boundary  is $\text{SNR}_{\min}=5$\,dB.
We repeat the simulations for $\eta=2$ (small cells) to intensify the effect of interference, especially copilot interference, received from the users in adjacent cells. We normalize the SNR such that the SNR before beamforming  for a user close to the BS remains the same in both scenarios.
%

\begin{figure}[t]
\centering
\subfloat[\label{PR3_geom}]{ \includegraphics[width=2.9cm]{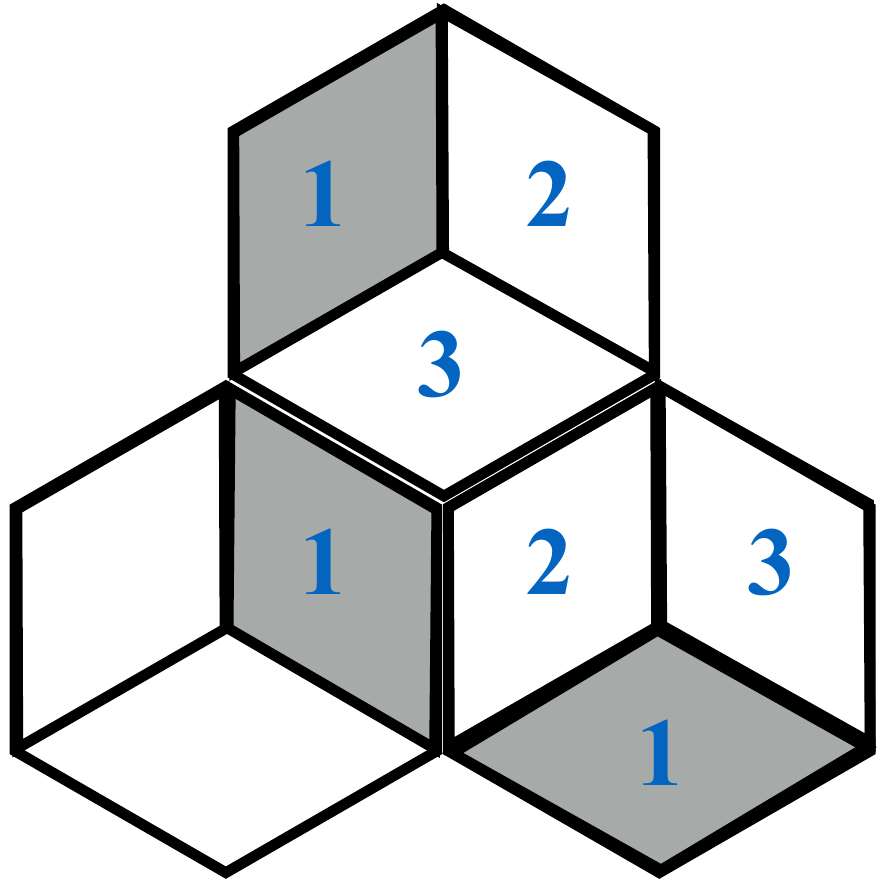} }
\hspace{1cm}
\subfloat[\label{PR1_geom}]{\includegraphics[width=2.9cm]{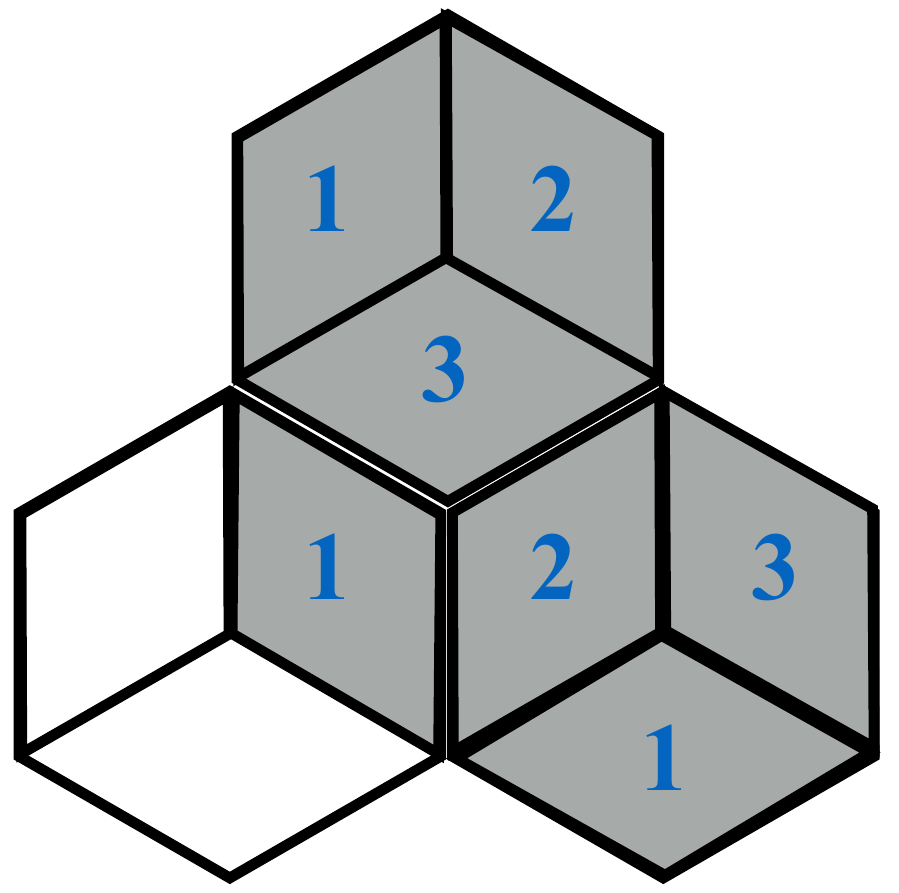} }
\caption{Pilot distribution for a  system with PR3 (a) and PR1 (b), with copilot sectors  highlighted in gray.}
\label{fig:pilot_geom}
\end{figure}

We assume that each hexagonal cell is divided into $3$ sectors as illustrated in Fig.\,\ref{fig:pilot_geom}. 
The BS uses a ULA with $M$ antennas to serve the users inside each sector, thus, the whole BS transmitter consists of $3$ ULAs (one per sector).
The ULAs are well isolated in the RF domain such that each ULA only receives the signal of the users lying in its 120\,deg angular span $[-\theta_{\max}, \theta_{\max}]$ with $\theta_{\max}=60$ degrees.

\subsection{Scattering Model}

We consider a one-ring scattering model for the user signal, where the transmitted signal from a user in the UL is reflected by a ring a scatterers located around the user with a radius of $R_\text{one-ring}=150$\,meters. We assume that all the scatterers contribute equally in terms of scattering power to the channel vector of the user observed at the BS. Thus, all the users have an equal delay-span of $\frac{2R_\text{one-ring}}{c_0}=1$\,$\mu$s but different angular 
spreads depending on their distance from the BS. 

\subsection{Physical Channel Model and OFDM Parameters}

We use a physical channel model similar to LTE (Long-Term Evolution) as in \cite{Marzetta-TWC10}. We consider a 
slot of duration $T_s = 0.532$\,ms and decide arbitrarily to send $7$ OFDM symbols over each slot, thus, each OFDM symbol has a total 
duration of $\frac{T_s}{7} = 76$\,$\mu$s and an effective duration $T_u = \frac{T_s}{7} - \Delta \tau_{\max} = 66$\,$\mu$s after removing the CP, 
corresponding to a frequency spacing of $\Delta f = \frac{1}{T_u} = 15$\,KHz between subcarriers.
We take a bandwidth of $W = 2$\,MHz with a frequency guard-band of $80$\,KHz, thus, the total number of subcarriers in each ODFM 
symbol is given by $N = \frac{1.92 \text{ MHz}}{15 \text{ KHz}} = 128$.

Assuming a coherence bandwidth of $\Delta f_c = 150$\,KHz, the number of subcarriers in each coherence sub-block (see Fig.\,\ref{pilot_structure}) is $D^\text{c}_\text{OFDM}= \Delta f_c T_u\approx 10$. Thus, the wideband channel matrix of $10$ users can be simultaneously learned over an individual training OFDM symbol, which  enforces a subcarrier sampling ratio $\frac{1}{10}$, i.e., we can  sample only $n = \lfloor \frac{N}{10} \rfloor = 12$ out of $N=128$ subcarriers of an OFDM symbol.
We devote $3$ OFDM symbols to channel estimation, where we are able to learn the channel matrix and, hence, serve up to $D_\text{p}=30$ users on a single TDD slot, consistently with the LTE-TDD standard.

{We simulate a sectorized cellular system with each cell consisting of $3$ sectors numbered $\{1,2,3\}$ as illustrated 
in Fig.\,\ref{fig:pilot_geom}. We consider a system with a \textit{Pilot Reuse} 3 (PR3) as illustrated in Fig.\,\ref{PR3_geom}, in which the 
set of $D^\text{c}_\text{p}= 30$ orthogonal pilots are shared among $3$ sectors such that sectors with similar numbers use identical set of pilots consisting of $10$ mutually orthogonal pilot sequences (i.e.,  $K=10$ served users in each sector), and groups of 3 adjacent sectors with different indices (1,2,3) use collectively 
all the 30 orthogonal pilot sequences.  We also consider a system with a \textit{Pilot Reuse} 1 (PR1) as illustrated in Fig.\,\ref{PR1_geom}, 
in which all the $30$ orthogonal pilot sequences are simultaneously used in all the sectors, thus, each sector can serve up to $K=30$ users.}

\begin{figure}[t]
\centering
\subfloat[\label{cls_pr1_1}]{%
\includegraphics[width=0.32\textwidth]{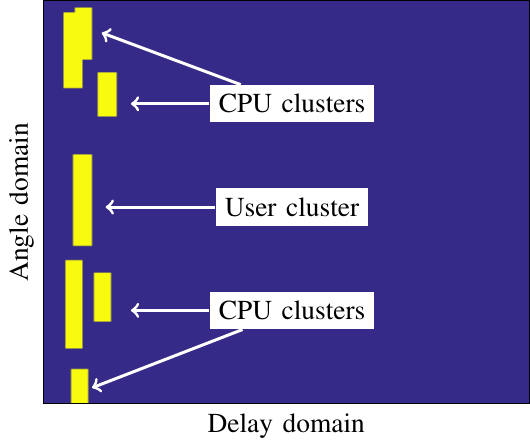}
}
\subfloat[\label{cls_pr1_2}]{%
\includegraphics[width=0.32\textwidth]{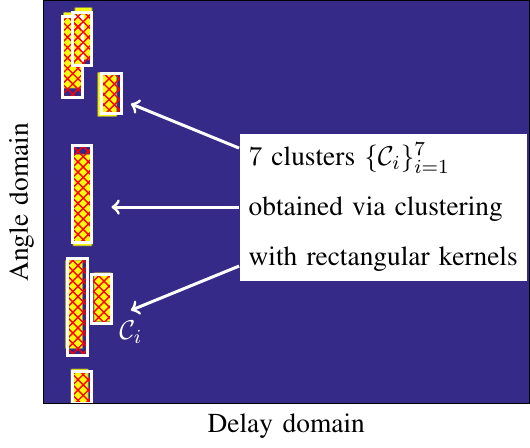}
}
\subfloat[\label{cls_pr1_3}]{%
\includegraphics[width=0.32\textwidth]{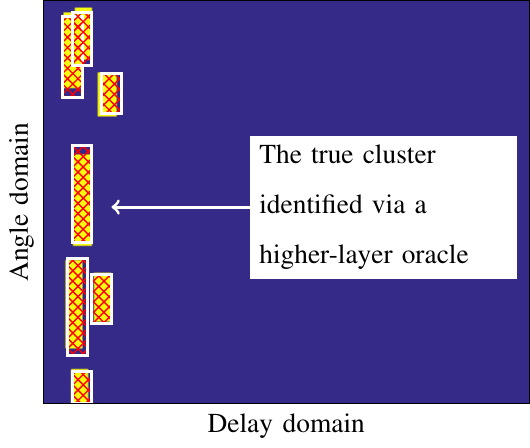}
}

\caption{Illustration of the supervised clustering in one-ring model. Fig.\,(a) illustrates the estimated angle-delay PSF, where it is not initially known which cluster belongs to the desired user.  Fig.\,(b) denotes a simple clustering of the estimated PSF into $7$ rectangular kernels ($1$ for the user and $6$ for its \cop s). Fig.\,(c) illustrates the identification of the true signal cluster via a higher-layer communication protocol.}
\label{sim_sup_cluster}
\end{figure}

\subsection{Clustering Algorithm}\label{sim_clst}
Since in PR3 illustrated in Fig.\,\ref{PR3_geom} users and their \cop s are well-separated in the delay domain, we apply an unsupervised clustering in the delay domain as in Fig.\,\ref{fig:clst_unsup} with a delay threshold 
$\tau_0=\frac{R_\text{cell}}{c_0}$. 
In particular, we set $\gammas(\theta,\tau)= \gamma(\theta, \tau) {\bf 1}_{\{(\theta,\tau) \in \clC^{\text{PR3}}_\ttd\}}$ and $\gammai(\theta,\tau)= \gamma(\theta, \tau) {\bf 1}_{\{(\theta,\tau) \notin \clC^{\text{PR3}}_\ttd\}}$, where $\clC^{\text{PR3}}_\ttd$ in the desired signal cluster given by
$\clC^{\text{PR3}}_\ttd=\Big \{(\theta,\tau): \tau\leq \tau_0 \Big \}$. 
{For PR1, the users and their \cop s are not generally separable in the delay domain (see, e.g., Fig.\,\ref{cls_pr1_1}). Here, we need to apply a supervised clustering algorithm to identify the desired user cluster. In the one-ring scattering model we consider for the simulations, the PSF of each user consists of a single angle-delay cluster (bubble). Since the number of \cop s is at most $6$, we cluster the estimated PSF into $7$ rectangular-shaped clusters illustrated in Fig.\,\ref{cls_pr1_2}. This separates approximately the clusters corresponding to the user and its \cop s but does not specify yet which cluster corresponds to the user. To identify the user cluster, we use the ``oracle'' provided from a higher communication layer with the following scheme. 
After receiving the noisy contaminated channel sketch $\bbx_s$ during a pilot transmission slot, we obtain estimates $\widehat{\bbh}_s^i$, $i=1,2,\dots, 7$, of the decontaminated channel vector of the user by treating the $i$-th cluster as the true signal cluster and the rest as \cop\ clusters, and applying 
our proposed channel interpolation algorithm in Section \ref{sec:masking}. During the data transmission phase, after receiving the whole ODFM symbol, 
we decode the received data  by beamforming along the columns of the channel matrix corresponding to $\widehat{\bbh}_s^i$, $i=1,\dots, 7$, 
once at a time, where we assume that there is a higher-layer oracle that selects the $\widehat{\bbh}_s^i$ and the corresponding cluster that results 
in a successful decoding of the user data as illustrated in Fig.\,\ref{cls_pr1_3}.}

%


\subsection{Uplink Pilot Decontamination}

For simulations, we focus on pilot decontamination in an UL scenario, where we focus on the users belonging to  Sector $1$ as in Fig.\,\ref{fig:pilot_geom}. The dominant copilot interference for each one of those users in the UL  comes from its $2$ nearest neighbor \cop s in PR3 and from its $6$ nearest neighbor \cop s in PR1. For each user, the BS learns the superposition of the channel vector of that user plus those of its \cop s. Note that due to the orthogonality of the pilots, during the UL training phase, there is only copilot interference but no interference from the other users. During the data transmission phase (UL or DL), however, there is a coherent interference from \cop s and a noncoherent interference from all the other users. Notice also that \cop s coming from 
non-nearest neighbor copilot sectors are received at significantly lower power and at larger delays. Such signals are not guaranteed to be eliminated by the proposed method since the OFDM model fails due to inter-block interference
(MPCs whose delays go beyond the CP interval). Nevertheless, the effect of non-nearest neighbors copilot contamination is
 very small.

\subsection{Antenna Sampling and Wideband Pilot Sketches}
We consider an antenna sampling ratio of $0.25$, where over each OFDM training symbol 
only $m = 0.25 M$ of the whole number of antennas $M$ are sampled. We assume that the sampling pattern is completely 
random and changes i.i.d. over time. 
We take a window of size $\sfw = 100$ of sketches across $\sfw$ time slots to estimate the channel geometry of each user, where we assume that  the channel matrices inside the window are i.i.d. since they belong to different slots (coherence times). As the whole observation takes  $\sfw T_s = 50$\,ms, we can safely assume that the channel geometry remains invariant over the whole window.

\subsection{Pilot Decontamination, Channel Interpolation, and Beamforming}

After estimating the PSF of all the users, we apply the clustering algorithm explained in Section \ref{sim_clst} and the masking technique as in \eqref{mask_M} to obtain the signal mask $\clM^{\bbh}$ and the interference mask $\clM^{\bbi}$ for each user, which we use for the rest of the time. 
We next simulate the communication phase, where each CB consists of a training phase to estimate the instantaneous channel vectors of the user and a data-transmission phase to send data to these users via spatial beamforming.   In each training slot $s$, after receiving a sketch of the channel vector of each user, we apply the low-complexity channel interpolation and decontamination algorithm in Section \ref{sec:masking} to estimate the full channel vector of the user. For simplicity of comparison with the contaminated case, we assume no antenna sampling is applied during a  training slot. 
We denote the decontaminated channel vectors of the $K$ users at the reference BS by $\{\widehat{\bbh}_{k}\}_{k=1}^{K}$ 
and the corresponding channel matrices by 
$\{\widehat{\bfH}_{k}\}_{k=1}^{K}$, where for simplicity we dropped the dependence on the data transmission slot $s$. 
We also assume that the noise power $\sigma^2$ in each antenna is available at the BS.

In PR3, we apply the MMSE beamforming for each user in the UL, where the normalized beamforming vector for a user $u \in [K]$ at 
subcarrier $\omega$ is given by by $\bfg_{k}[\omega]=\frac{\bfv_{k}[\omega]}{\|\bfv_{k}[\omega]\|}$, where
\begin{align}\label{mmse_BF}
\bfv_{k}[\omega]=\Big ( \sigma^2 \bfI_M + \sum_{k'} \widehat{\bfh}_{k'}[\omega] {\widehat{\bfh}_{k'}[\omega]}^\herm \Big )^{-1} \widehat{\bfh}_{k}[\omega],
\end{align}
where $\widehat{\bfh}_{k}[\omega]$ denotes the decontaminated and interpolated channel vector of the user $k$ at subcarrier $\omega$.
Due to the sectorization, the BS in Sector 1 not only receives the pilot signal of its users but can also listen to the pilot signal of the users in adjacent sectors since they are using disjoint set of pilots. Thus, the summation over $k'$ in \eqref{mmse_BF} is taken over all the users inside the sector as well as 
the users in adjacent sectors. For PR1, we use a simple {\em conjugate beamforming} \cite{Marzetta-TWC10} given by $\bfg_{k}[\omega]=\frac{\widehat{\bfh}_{k}[\omega]}{\|\widehat{\bfh}_{k}[\omega]\|}$.
We compare the performance of our method with the case where no pilot decontamination is applied. We assume that in such a case also $1$ out of $10$ of columns of channel matrix of each user is observed during a training slot. We apply DFT interpolation to interpolate the unobserved columns of the channel matrix of each user and repeat similar steps, as in the decontaminated case, to design the beamforming vectors. We define the SINR of the channel of the user $k \in [K]$ at subcarrier $\omega$ by 
\begin{align}
\sinr_{k}[\omega]=\frac{\Big|\bfg_{k}[\omega]^\herm \bfh_{k}[\omega]\Big |^2 }{\sigma^2 + \sum_{k' \neq k}\Big|\bfg_{k}[\omega]^\herm \bfh_{k'}[\omega]\Big |^2},
\end{align}
where the summation is taken over all the users $k'\neq k$ inside the sector as well as all other users in adjacent cells who lie in the angular 
span of the ULA of Sector $1$ and create interference. Assuming perfect channel state information after beamforming, the \textit{instantaneous spectral efficiency} of user $k$ is given by $R_{k} = \frac{1}{N} \sum_{\omega \in [N]} \log_2(1+\sinr_{k}[\omega])$. 
We denote the achievable sum-rate in bit/s/Hz of all the users in Sector $1$ by $R = \sum_{k \in [K]} R_{k}$, 
where $R$ is a random variable depending on the instantaneous realizations of the channel 
vectors of all the users. 

\begin{figure*}[t]
\centering

\subfloat[\label{pilot_sharing_beta3_avg}]{%
\includegraphics[width=0.4\textwidth]{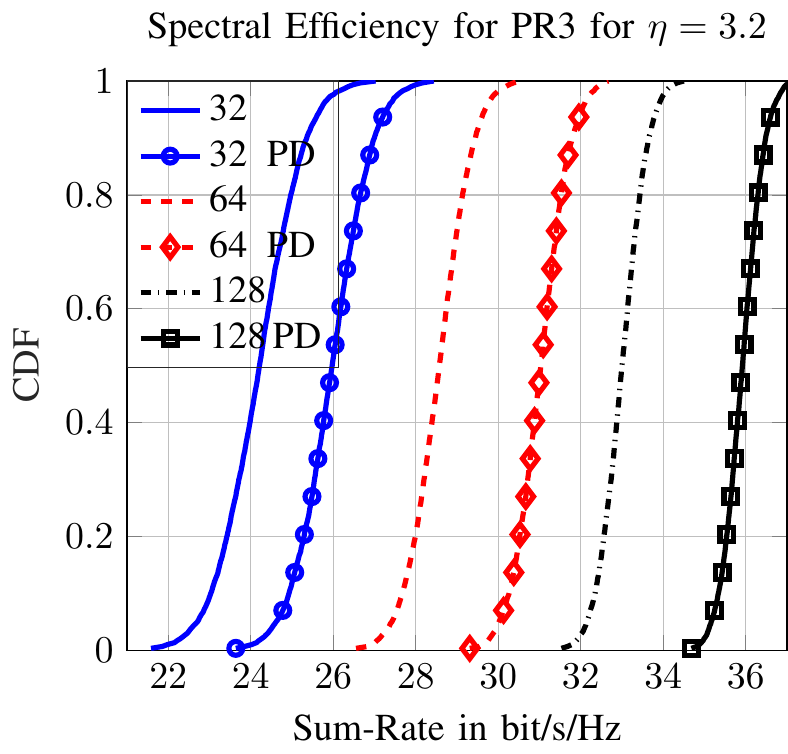}
}
\hfill
\subfloat[\label{pilot_sharing_beta2_avg}]{%
\includegraphics[width=0.4\textwidth]{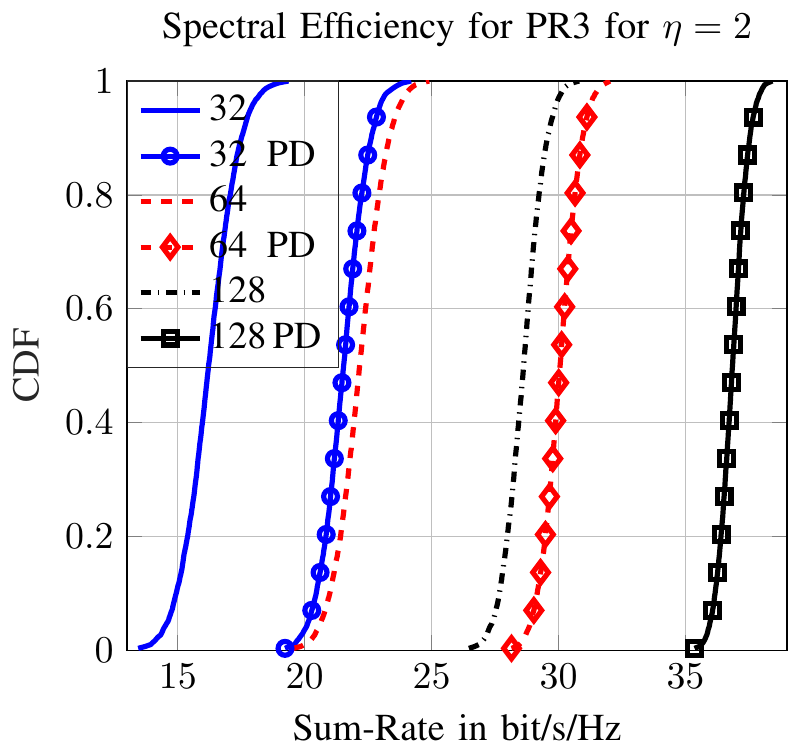}
}

\caption{CDF of Spectral Efficiency (Sum-Rate) in bit/s/Hz for different number of BS antennas $M\in \{32, 64, 128\}$ for PR3 with $\eta=3.2$  (a) and $\eta=2$  (b). The curves with marks and with the legend ``PD'' illustrate the CDF after pilot decontamination. All the plots are averaged over $N_\text{geom}=30$ random user locations in the system.}
\label{fig:simulation_PR3}
\end{figure*}

\subsection{Achievable Performance and Comparison with the State of the Art}\label{sec:performance}

\textit{1) PR3:} Fig.\,\ref{fig:simulation_PR3} illustrates the \textit{Cumulative Distribution Function} (CDF) $F_R(r)$ of the achievable spectral efficiency $R$ in bit/s/Hz before and after pilot decontamination. We average the CDFs over $N_\text{geom}=30$ independent  realizations of the geometry of users across the system.
 For each geometry realization, we run the simulations for different number of BS antennas $M \in \{32,64,128\}$. 
We also consider two different scenarios for two different power-loss exponents $\eta \in \{3.2, 2\}$ as explained in Section \ref{power_model}. 
It is seen that for $\eta=3.2$ and for practical numbers of BS antennas $M \in \{32,64,128\}$ pilot decontamination improves the spectral efficiency by $10\%-20\%$ for PR3, where the resulting gain increases by increasing the number of BS antennas $M$.  For $\eta=2$, on the other hand, our proposed  scheme results in a dramatic gain in spectral efficiency.

\input{sim_PR1_smooth.tex}

{
\noindent \textit{2) PR1:} We repeat the simulations for PR1. In this case, to pinpoint the effect of pilot contamination, rather than calculating the sum-rate averaged over all the random locations of the users, we focus on an ``edge''  user randomly located on the cell boundary. 
We expect that the spectral efficiency of such a user be affected considerably by the pilot contamination from the neighboring  \cop s.  
As illustrated in Fig.\,\ref{sim_sup_cluster}, pilot decontamination for edge users in PR1 requires a supervised clustering 
in the angle-delay domain.  Fig.\,\ref{fig:simulation_PR1} illustrates the simulation results. We compare the performance of our algorithm with the 
one proposed in \cite{chen2016pilot}. In \cite{chen2016pilot}, the  support of the MPCs of the contaminated channel vector of the user is estimated 
by devoting all the subcarriers in an OFDM symbol to an individual user and projecting the whole channel matrix in the 2D FFT basis. 
The support of the desired user is identified and separated from that of its \cop s by taking the intersection of the support 
obtained over several slots, where in each slot the pilots are shuffled such that the desired user collides with different \cop s at each slot. The rationale behind this idea is that in this way the support of the MPCs of the desired user remains constant over the sequence of slots, while
that of the \cop s changes from slot to slot. Therefore, taking the intersection of the estimated supports over the slots should yield the MPCs of the desired user.
However, in doing so, the intersection will also exclude the MPCs of the desired user that over the sequence of slots experience a deep fade, since these will be missed on some slots, and therefore will not be contained in the intersection. As a matter of fact, with time-selective fading as in our realistic setting, 
we could verify that the method of \cite{chen2016pilot} dramatically underestimates the MPCs of the desired user. 

In contrast, our proposed subspace estimation with supervised clustering is much robust to small-scale fading variations, 
performs much better in the presence of overlapping clusters (e.g., when a user and it \cop s have common clusters), 
and does not require pilot shuffling  among the neighboring cells. 
Also, compared with \cite{chen2016pilot}, in our proposed scheme only a fraction (e.g., $\frac{1}{10}$) of the subcarriers in an OFDM symbol are devoted 
as pilot to each user, so the channel state of several users (e.g., $10$ users) can be simultaneously estimated over an individual OFDM symbol, 
thus, much better multiplexing gain. Notice that using the whole set of $N$ OFDM subcarriers for UL pilots is essential to the method of 
 \cite{chen2016pilot} since otherwise there is not enough resolution in the delay domain. This is because  \cite{chen2016pilot} makes use of simple linear projections, while our scheme estimates the PSF using the advanced $l_{2,1}$-regularized least squares minimization described in Section \ref{sec:ch_estim}. 

Our simulation results in Fig.\,\ref{fig:simulation_PR1} consider the rate CDF of a single edge user randomly located near the 
cell boundary, thus, they do  not reflect the additional multiplexing gain resulting from using a reduced pilot dimension $n \ll N$. 
For the simulations, we assume that the total number of users is the same in both scenarios ($30$ users per sector), 
where similar to PR3 we average the achievable spectral efficiency of this specific edge user over $N_\text{geom}=30$ independent 
realizations of the geometry of all the users across the whole system. From Fig.\,\ref{fig:simulation_PR1}, it is seen that the gain 
in spectral efficiency obtained by our method is much more than the one proposed in \cite{chen2016pilot}. In particular, the resulting 
gain scales much better with the number of BS antennas.}

\section{Conclusions}
{
In this paper, we presented a novel scheme to eliminate the effect of pilot contamination on the performance of a massive MIMO wireless cellular system. 
We proposed a low-complexity algorithm that uses the pilot signal received from each user inside a window containing several time slots to obtain an estimate of the angle-delay power spread function (PSF) of each user contaminated channel vectors. We used the key idea, already exploited in various ways in the recent massive MIMO literature, that the channel vectors of each user consist of sparse MPCs in the angle-delay domain. We exploited this underlying sparsity to estimate the angle-delay PSF of each user by sampling only a small subset of antennas and, more importantly, by transmitting pilots  across only a subset of subcarriers compatible with LTE-TDD  without incurring any pilot overhead. We proposed clustering algorithms to decompose the estimated PSF of each user into its signal and copilot interference part. We exploited this decomposition to decontaminate the channel vector of each user in the next coherence blocks.  
Through Monte Carlo simulation, we demonstrated the effectiveness of the proposed pilot-decontamination scheme for practical scenarios with practical user geometries, reasonable number of BS antennas $M \in \{32,64,128, 256\}$, and realistic fading channel statistics as in \cite{Marzetta-TWC10}. We also compared our proposed method with the competitive scheme \cite{chen2016pilot} and illustrated that our method provides much better performance in terms of multiplexing gain, pilot decontamination efficiency, and scaling performance with the number of BS antennas. 
}

\appendices

\section{Low-complexity Interpolation using ADMM}\label{sec:admm}

Consider the following cost function as in \eqref{c_PQ}:
\begin{align}\label{c_PQ2}
c(\bfP,\bfQ)&=\frac{1}{2} \|\bfX_s - {\bfS_s^\tta} (\bfP+\bfQ) {{\bfS_s^\ttf}}^\herm\|^2 + \ind_{\clM^\bbh}(\fovs(\bfP))+ \ind_{\clM^\bbi}(\fovs(\bfQ)).
\end{align} 
In this section, we assume that the convex regularizers $\ind_{\clM^\bbh}$ and $\ind_{\clM^\bbi}$ are the indicator functions of ${\clM^\bbh}$ and ${\clM^\bbi}$ defined similarly to \eqref{ind_def}. We first introduce the auxiliary variables $\bfP_\ttf=\fovs(\bfP)$ and $\bfQ_\ttf=\fovs(\bfQ)$ of dimension $G^\theta\times G^\tau$ and define
\begin{align}\label{c_PQ3}
c(\bfP,\bfQ, \bfP_\ttf, \bfQ_\ttf)&=\frac{1}{2} \|\bfX_s - {\bfS_s^\tta} (\bfP+\bfQ) {{\bfS_s^\ttf}}^\herm\|^2 + \ind_{\clM^\bbh}(\bfP_\ttf)+ \ind_{\clM^\bbi}(\bfQ_\ttf).
\end{align} 
Thus, minimizing $c(\bfP,\bfQ)$ in \eqref{c_PQ2} can be equivalently written as minimizing $c(\bfP,\bfQ, \bfP_\ttf, \bfQ_\ttf)$ under the additional linear constraints $\bfP_\ttf=\fovs(\bfP), \bfQ_\ttf=\fovs(\bfQ)$, which is still a convex optimization problem. We use \textit{Alternating Direction Method of Multipliers} (ADMM) to solve this optimization problem. We introduce the Lagrange variables $\Lambdam_\ttp$ and $\Lambdam_\ttq$ of dimension $G^\theta\times G^\tau$ and the augmented Lagrangian function 
\begin{align}
\scrL&=\frac{1}{2}\|\bfX_s - {\bfS_s^\tta} (\bfP+\bfQ)  {{\bfS_s^\ttf}}^\herm\|^2 + \ind_{\clM^\bbh}(\bfP_\ttf)+ \ind_{\clM^\bbi}(\bfQ_\ttf)\nonumber\\
&+\inpr{\Lambdam_\ttp}{\bfP_\ttf-\fovs(\bfP)} + \inpr{\Lambdam_\ttq}{\bfQ_\ttf-\fovs(\bfQ)}\nonumber\\
&+ \frac{\upsilon}{2} \|\bfP_\ttf-\fovs(\bfP)\|^2 + \frac{\upsilon}{2} \|\bfQ_\ttf-\fovs(\bfQ)\|^2,\label{aug_lag}
\end{align}
where $\upsilon \in \bR_+$ is the ADMM parameter to be set. The ADMM iteration can be written as follows:
\begin{align}
(\bfP^{k+1}, \bfQ^{k+1})&=\argmin_{\bfP,\bfQ} \scrL (\bfP,\bfQ,\bfP_\ttf^k,\bfQ_\ttf^k, \Lambdam_\ttp^k, \Lambdam_\ttq^k),\label{pq_iter}\\
(\bfP_\ttf^{k+1}, \bfQ_\ttf^{k+1})&=\argmin_{\bfP_\ttf,\bfQ_\ttf} \scrL (\bfP^{k+1},\bfQ^{k+1},\bfP_\ttf,\bfQ_\ttf, \Lambdam_\ttp^k, \Lambdam_\ttq^k),\label{pqf_iter}\\
\Lambdam_\ttp^{k+1}&=\Lambdam_\ttp^k + \upsilon(\bfP_\ttf^{k+1}-\fovs(\bfP^{k+1})), \label{lam_p}\\
\Lambdam_\ttq^{k+1}&=\Lambdam_\ttq^k + \upsilon(\bfQ_\ttf^{k+1}-\fovs(\bfQ^{k+1})), \label{lam_q}
\end{align}

\noindent
\colorbox{gray!40}{ 
\begin{minipage}{0.205\textwidth}
\noindent \hspace{-2mm}\textit{Updating $\bfP^{k+1},\bfQ^{k+1}$:}
\end{minipage}
}
%
%
\hspace{-1mm}Using the vectorization and denoting by $\bbp=\vec(\bfP)$, $\bbq=\vec(\bfQ)$, $\bbx_s=\vec(\bfX_s)$, $\bfP_\ttt=\fovs^{-1}(\bfP_\ttf^k + \frac{1}{\upsilon} \Lambdam^k_\ttp)$, $\bfQ_\ttt=\fovs^{-1}(\bfQ^k_\ttf + \frac{1}{\upsilon} \Lambdam^k_\ttp)$, $\bbp_\ttt=\vec(\bfP_\ttt(\range{1}{M}, \range{1}{N}))$, $\bbq_\ttt=\vec(\bfQ_\ttt(\range{1}{M}, \range{1}{N}))$, we can write \eqref{pq_iter} as the following cost function to be minimized for $MN\times 1$ vectors $\bbp$ and $\bbq$:
\begin{align}\label{admm_vec_version}
\frac{1}{2}\|\bbx_s - \bS_s (\bbp+\bbq)\|^2+ \frac{\upsilon}{2} \|\bbp-\bbp_\ttt\|^2 + \frac{\upsilon}{2} \|\bbq- \bbq_\ttt\|^2,
\end{align}
where $\bS_s=\bfS_s^\ttf \otimes \bfS_s^\tta$ denotes the sampling operator at slot $s$. The optimal solution of \eqref{admm_vec_version} is given by 
\begin{align}
\bbp^{k+1}&=\bbp_\ttt + \frac{\check{\bbx}_s}{\upsilon}  + \scrS_s\big(2 \scrS_s + \upsilon\scrI\big )^{-1} (\bbp_\ttt + \bbq_\ttt+ \frac{2\check{\bbx}_s}{\upsilon}),\label{p_up}\\
\bbq^{k+1}&=\bbq_\ttt + \frac{\check{\bbx}_s}{\upsilon}  + \scrS_s\big(2 \scrS_s + \upsilon\scrI\big )^{-1} (\bbp_\ttt + \bbq_\ttt+ \frac{2\check{\bbx}_s}{\upsilon})\label{q_up},
\end{align}
where $\check{\bbx}_s=\bS_s^\herm \bbx_s$, $\scrS_s=\bS_s^\herm \bS_s$, and where $\scrI=\bfI_{MN}$ denotes the identity matrix of order $MN$. Since in this paper we always use 0-1 antenna and frequency sampling matrices, \eqref{p_up} and \eqref{q_up} can be further simplified. Using the properties of the $\vec$ operator, we have that
\begin{align}
\scrS_s=\bS_s^\herm \bS_s=({\bfS_s^\ttf}^\herm\otimes {\bfS_s^\tta}^\herm)({\bfS_s^\ttf}\otimes {\bfS_s^\tta})= ({\bfS_s^\ttf}^\herm {\bfS_s^\ttf})\otimes ({\bfS_s^\tta}^\herm {\bfS_s^\tta}).
\end{align}
Note that, due to 0-1 sampling, ${\bfS_s^\ttf}^\herm {\bfS_s^\ttf}$ and ${\bfS_s^\tta}^\herm {\bfS_s^\tta}$ are diagonal matrices of dimension $M\times M$ and $N\times N$ with 1s in the diagonal elements corresponding to the index sets $\clI_s^\ttf$ and $\clI_s^\tta$ and $0$ elsewhere, where $\clI_s^\tta \subseteq[M]$ and $\clI_s^\ttf\subseteq [N]$ denote the indices of antennas and subcarriers sampled at slot $s$ as explained in Section \ref{chmod_subsamp}. This implies that $\scrS_s$ is a 0-1 diagonal matrix of dimension $MN\times MN$ where the locations of 1s in the diagonal is given as in \eqref{I_set} by 
\begin{align}
\clI_s:=\{M(c_s^\ttf -1)+c_s^\tta: \ \ c_s^\ttf \in \clI_s^\ttf,\ c_s^\tta \in \clI_s^\tta\} \subseteq [MN].
\end{align}
As a result, the matrix $\scrS_s\big(2 \scrS_s + \upsilon\scrI\big )^{-1}$ in \eqref{p_up} and \eqref{q_up} is a diagonal matrix with a value $\frac{1}{\upsilon+2}$ at the diagonal elements belonging to $\clI_s$ and $0$ elsewhere. Moreover, no matrix-vector multiplication is needed for computing $\check{\bbx}_s=\bS_s^\herm \bbx_s$ since $\check{\bbx}_s$ is simply given by an $MN\times 1$ vector that contains the components of $\bbx_s$ in the indices corresponding to $\clI_s$ and is $0$ elsewhere. This implies that $\bbp^{k+1}$ and $\bbq^{k+1}$ can be easily computed from \eqref{p_up} and \eqref{q_up}, from which we obtain $\bfP^{k+1}$ and $\bfQ^{k+1}$ via inverse $\vec$ operation.
The whole computational complexity of this step comes from calculating $\bbp_\ttt$ and $\bbq_\ttt$, which requires $O(G \log_2(G))$ operations where $G=G^\theta G^\tau$ denotes the grid size as  before.

\vspace{1mm}
\noindent
\colorbox{gray!40}{ 
\begin{minipage}{0.205\textwidth}
\noindent \hspace{-2mm}\textit{Updating $\bfP_\ttf^{k+1},\bfQ_\ttf^{k+1}$}:
\end{minipage}
}
We first derive the update equation for $\bfP_\ttf^{k+1}$ in \eqref{pqf_iter}. To find $\bfP_\ttf^{k+1}$, we need to optimize the following function with respect to $\bfP_\ttf$:
\begin{align}\label{pf_upd_res}
\frac{\upsilon}{2} \|\bfP_\ttf + \frac{1}{\upsilon} \Lambdam^k_\ttp -\fovs(\bfP^{k+1})\|^2 + \bI_{\clM^\bbh}(\bfP_\ttf).
\end{align}
The optimal solution of \eqref{pf_upd_res} is given by setting $\bfP^{k+1}_\ttf$ equal to $\fovs(\bfP^{k+1}) - \frac{1}{\upsilon}\Lambdam^k_\ttp$ at those elements belonging to the mask $\clM^\bbh$ while setting the remaining components equal to zero. Similarly, $\bfQ^{k+1}_\ttf$ is given by $\fovs(\bfQ^{k+1}) - \frac{1}{\upsilon}\Lambdam^k_\ttq$ over the mask $\clM^\bbi$ and zero elsewhere. The whole computational complexity of this step is also $O(G \log_2(G))$ for computing $\fovs(\bfP^{k+1})$ and $\fovs(\bfQ^{k+1})$. 
Overall, the computational complexity of each ADMM iteration is $O(G \log_2(G))$.

\begin{figure}[t]
\centering
\subfloat[2D DFT of received signal.\label{initial}]{%
\includegraphics[width=4.2cm]{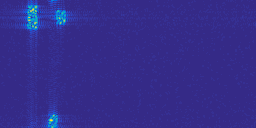}
}
\subfloat[After $1$ ADMM iteration.\label{iter-1}]{%
\includegraphics[width=4.2cm]{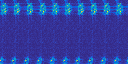}
}
\subfloat[After $2$ ADMM iterations.\label{iter-2}]{%
\includegraphics[width=4.2cm]{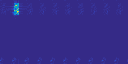}
}
\subfloat[After $3$ ADMM iterations.\label{iter-3}]{%
\includegraphics[width=4.2cm]{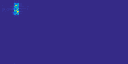}
}
\caption{Illustration of ADMM algorithm for a user in a system with PR3. Fig.\,(a) illustrates the 2D DFT of noisy contaminated channel vector of the user in the angle (vertical) and delay (horizontal) domain. The channel vector consists of  a desired signal cluster along with $2$ interfering copilot clusters. Fig.\,(b) show the 2D DFT after the first iteration, which also shows the appearance of aliasing pattern (with $10$ replicas) due to subsampling (by a factor $10$) in the frequency (subcarrier) domain. From Subfig.\,(c) and (d), it is seen that the algorithm reconstructs the true channel matrix quite fast. }
\label{fig:ADMM}
\end{figure}

\subsection{Simulation Results}
Fig.\,\ref{fig:ADMM} illustrates the performance our proposed ADMM algorithm in decontaminating/interpolating the channel vector of the user. For simulation, we consider a user inside a cellular system with PR3 as illustrated in Fig.\,\ref{fig:pilot_geom}. The Subfig.\,(a) in Fig.\,\ref{fig:ADMM} shows the 2D DFT of the received noisy and contaminated channel vector in the angle-delay domain. It is seen that the received signal contains a desired signal cluster with smaller propagation delay and two copilot clusters with larger delays. It is seen that our proposed masking technique in Section \ref{sec:masking} along with the ADMM implementation reconstructs the decontaminated channel matrix quite fast. Interestingly, in this example, one of the copilot clusters overlaps with the signal cluster in the angle domain, thus, the column-wise decontamination of the channel matrix, as proposed in Section \ref{sec:angle_separ}, will not be effective. 

\balance
{\footnotesize
\bibliographystyle{IEEEtran}
\bibliography{references2}
}

\end{document}